\documentclass[journal]{IEEEtran}
\usepackage{amsmath,graphicx,epstopdf}
\usepackage{epsfig}
\usepackage{citesort}
\usepackage{multirow,bigstrut}
\usepackage{amssymb}
\usepackage{color}
\usepackage{url}
\usepackage{booktabs}
\usepackage{diagbox}
\usepackage{subfigure}
\usepackage{overpic}
\begin{document}
\title{
Combining Progressive Rethinking 
and Collaborative Learning: 
A Deep Framework for In-Loop Filtering 
}

\author{Dezhao Wang, \textit{Student Member, IEEE}, Sifeng Xia, Wenhan Yang, \textit{Member, IEEE}, \\Jiaying Liu, \textit{Senior Member, IEEE}
\thanks{			
			This work was supported
			in part by the National Key Research and Development Program of China under Grant No. 2018AAA0102702,
			in part by the Fundamental Research Funds for the Central Universities, 
			in part by the National Natural Science Foundation of China under Contract No.61772043, 
			The associate editor coordinating the review of this manuscript and approving it for publication was Dr. Adrian Munteanu.
			(Corresponding author: Jiaying Liu.)
		}
		\thanks{
			Dezhao Wang, Sifeng Xia, Wenhan Yang and Jiaying Liu is with the Wangxuan Institute of Computer Technology, Peking University, Beijing, 100080, China. (e-mail: wangdz@pku.edu.cn, xsfatpku@pku.edu.cn, yangwenhan@pku.edu.cn ,liujiaying@pku.edu.cn)
		}

}

\markboth{IEEE TRANSACTIONS ON IMAGE PROCESSING}%
{Shell \MakeLowercase{\textit{et al.}}: Bare Demo of IEEEtran.cls for IEEE Journals}

\maketitle

\begin{abstract}
In this paper, we aim to address issues of (1) joint spatial-temporal modeling and (2) side information injection for deep-learning based in-loop filter. For (1), we design a deep network with both progressive rethinking and collaborative learning mechanisms to improve quality of the reconstructed intra-frames and inter-frames, respectively. For intra coding, a Progressive Rethinking Network (PRN) is designed to simulate the human decision mechanism for effective spatial modeling. Our designed block introduces an additional inter-block connection to bypass a high-dimensional informative feature before the bottleneck module across blocks to review the complete past memorized experiences and \textit{rethinks progressively}. For inter coding, the current reconstructed frame interacts with reference frames (peak quality frame and the nearest adjacent frame) \textit{collaboratively} at the feature level. For (2), we extract both intra-frame and inter-frame side information for better context modeling. A coarse-to-fine partition map based on HEVC partition trees is built as the intra-frame side information. Furthermore, the warped features of the reference frames are offered as the inter-frame side information. Our PRN with intra-frame side information provides 9.0\% BD-rate reduction on average compared to HEVC baseline under All-intra (AI) configuration. While under Low-Delay B (LDB), Low-Delay P (LDP) and Random Access (RA) configuration, our PRN with inter-frame side information provides 9.0\%, 10.6\% and 8.0\% BD-rate reduction on average respectively. Our project webpage is \textcolor{magenta}{\url{https://dezhao-wang.github.io/PRN-v2/}}.
\end{abstract}


\IEEEpeerreviewmaketitle

\section{Introduction}
\label{intro}
\IEEEPARstart{L}{ossy} video compression is widely applied due to its effectiveness in bit-rate saving and critical visual information preservation.
However, these two goals are contradictory and it is non-trivial to optimized them jointly.
Modern video compression standards such as High Efficient Video Coding (HEVC) \cite{HEVC} still suffer from various kinds of degradation for the sake of block-wise processing and quantization.

To remove these artifacts, an in-loop filter module consisting of 
Deblocking Filter (DF)~\cite{df} 
and Sample Adaptive Offset~(SAO)~\cite{sao}
is applied to suppress blocking and ringing artifacts. 
The in-loop filter not only effectively enhances the quality of the reconstructed frames and further benefits the subsequent inter-coding procedure via providing high quality reference frames.
Lots of efforts are put into this field, improving the quality of the reconstructed frames in the coding loop, and a series of works are proposed based on handcrafted filters~\cite{texture_preservation},
Markov random filed~\cite{SunC07},
nonlocal filters~\cite{zhang12},
low-rank minimization~\cite{dcc_low_rank}, \textit{etc}.
However, these methods built on shallow models offer limited performance.

In recent years, deep learning brings in new progresses in related fields, firstly image and video restorations for low-level visions, and leads to impressive performance gains.
A series of milestone network architectures and basic blocks are proposed, \textit{e.g.} Super-Resolution Convolutional Neural Network (SRCNN)~\cite{srcnn},
Very Deep Super-Resolution network (VDSR)~\cite{vdsr},
Denoising Convolutional Neural Network (DnCNN)~\cite{dncnn},
and Dual-domain Multi-scale Convolutional Neural Network (DMCNN)~\cite{dmcnn} for compression artifacts removal, \textit{etc.}
The latest methods become more advanced,
and usually make use of the power of residual learning, dense connections, or their combinations, in a cascaded or recurrent manner.
For example, Lim \textit{et al.}~\cite{edsr} proposed to cascade multiple residual blocks as Enhanced Deep Super-Resolution Network (EDSR).
Later on, Zhang~\textit{et al.}~\cite{rdn} embedded the dense connections~\cite{densenet} into a residual network~\cite{resnet}. 
Inspired by the recent development of these works, many deep-learning based in-loop filtering methods and post-processing methods are proposed~\cite{icip18_mask,drn,MIF,vrcnn}, from the simplest cascaded CNN~\cite{vrcnn} to the combination of residual learning and dense connections~\cite{drn}.

Besides the network architecture evolution, video coding scenario also provides rich context side information to improve the quality of the reconstructed frames.
For example, the partition structure of the coding process implicitly reveals the structural complexity of local regions and indicates the relative quality loss after the compression.
For convenience, based on whether inferred with adjacent frames, we classify the side informations into intra-frame and inter-frame side information. 
Inspired by Kalman Filter, Lu \textit{et al.}~\cite{kalman} proposed a Deep Kalman Filtering Network (DKFN) to take the extracted quantized prediction residual image from the codec as another input.
When it comes to the in-loop filter, there is also useful side information proposed in HEVC codecs.
For example, He \textit{et al.}~\cite{icip18_mask} proposed a post-processing network taking the partition mask, inferred based on the partition tree of HEVC, as the side information.
In~\cite{CPHER}, an EDSR-like network takes the unfiltered and prediction frames as side information and is trained with weight normalization.
For inter-frame side information, it is intuitive to make use of temporal redundancy to obtain useful information from the adjacent frames to benefit the processing of the current frame.
In~\cite{MIF}, the reference frames (the nearest adjacent frames or the peak quality frame) are warped by optical flow or designed motion compensation modules, and taken as another input to improve the quality of the current frame.

Although achieving significant performance improvements compared to previous works, these methods still have ignored issues from the perspectives of model design, coding context perception, and side information utilization.
\begin{itemize}
	\item At the model design level, the most popular network architectures~\cite{rdn,drn} for in-loop filtering and the low-level tasks combine the power of residual learning and dense connections by stacking several basic blocks.
	The channel dimensions of the output features across  blocks are usually compressed to make the output feature compact to prevent introducing too many parameters. However, this compression also leads to the information loss and limits the modeling capacity.
	\item In the video coding scenarios, the temporal modeling is quite different from that in video restoration/enhancement tasks from two aspects.
	First, the video frames might be reordered based on different coding configurations. Second, the quality of the reconstructed frames varies a lot.
	Previous works make use of temporal redundancies by taking the warped frames as input.
	This way does not exhaust the potential of modeling capacities, which is buried in the complex dependencies of video frames in the coding scenario.
	\item For side information utilization, some side information is not  considered closely with the coding context and its potential is not fully explored. For example, 
	the partition masks used in~\cite{icip18_mask} are only inferred from the leaf nodes of the partition tree. 
	In fact, the nodes on different levels of the partition tree can provide regional context information at different granularity.
\end{itemize}

In this paper, we aim to address the three issues mentioned above.
Specifically, we develop a deep network with both \textit{progressive rethinking} and \textit{collaborative learning} mechanisms to improve quality of the reconstructed intra-frames and inter-frames, respectively.
The progressive rethinking mechanism improves the modeling capacity of the in-loop filtering baseline network for both intra-frames and inter-frames.
Inspired by the human decision mechanism, a Progressive Rethinking Block (PRB) and its stacked Progressive Rethinking Network (PRN) are designed.
They are different from typical cascaded deep networks, where at the end of each basic block, the dimension size of the feature is reduced to generate the summarization of the past experiences.
Our PRN takes a \textit{Progressive Rethinking} manner.
The PRB introduces an additional inter-block connection to bypass a high-dimensional informative feature across blocks to review the complete past memorized experiences.
The \textit{Collaborative Learning Mechanism} tries to fully explore the potential of temporal modeling in the video coding scenario.
It acts like the collaboration of human being, where information is exchanged and refined progressively.
The current reconstructed frame interacts with the reference frames (peak quality frame and the nearest adjacent frame) progressively at the feature level. Therefore, they complement for each other's information deeply. Furthermore, novel intra-frame and inter-frame side informations are designed for a better context modeling. 
A coarse-to-fine partition map based on HEVC partition trees is built as the intra-frame side information.
Besides, the warped features of the reference frames are offered as the inter-frame side information.

This paper is an extension of our conference paper \cite{PRN}. Beyond single frame in-loop filtering, we further develop a Progressive Rethinking Recurrent Neural Network which utilizes temporal information to guide the restoration. To efficiently filter the current frame, we pick up two reference frames as an aid and share the information among these frames by a Collaborative Learning Mechanism, which further improves the coding performance. Moreover, we add extensive experiments and model analysis in this paper to show the effectiveness of our method and rationality of our model design.

In summary, our contributions are three-fold:
\begin{enumerate}
    \item We design a Progressive Rethinking Block based on residual learning and dense connection. An additional inter-block connection is proposed to compensate for the lost information caused by dimension compression, which improves the modeling capacity for in-loop filters of both intra-frames and inter-frames.
    \item We propose a Progressive Rethinking Recurrent Neural Network (PR-RNN) for collaborative learning to effectively utilize temporal redundancies in the video coding scenario.
    Motivated by the collaboration among human beings, we update the states of the current frame as well as reference frames synchronously by information sharing progressiveness.     
    \item We exploit context side informations from HEVC codecs to better adapt to the coding scenario. We extract Multi-scale Mean value of CU (MM-CU) maps from the partition tree to guide the network restoration. By fusing MM-CUs to the baseline network we establish our Progressive Rethinking Convolutional Neural Network (PR-CNN) as an effective single frame filter under All-Intra (AI) configuration.
\end{enumerate}
The remainder of the paper is organized as follows. In Section \ref{sec2}, we provide a brief review of related works. In Section \ref{sec3}, we introduce the methodology of our Progressive Rethinking Networks (PRN). Section \ref{sec4} provides the implementation. Experimental results are shown in Section \ref{sec5}. Finally we will make a conclusion in Section \ref{sec6}.

\section{Related Works}
\label{sec2}
\subsection{Deep Learning Based Video Coding}
Modern video coding standards like HEVC consist of multiple modules working together to compress the given videos. With the development of deep learning, researchers begin to utilize the strong non-linear mapping capability to substitute or enhance the original module in the codecs.

In \cite{ipfcn},  Li \textit{et al.} developed a fully-connected neural network for intra prediction (IPFCN). The IPFCN takes the neighbouring pixels as input to predict the current block pixel values. Hu \textit{et al.} proposed a Progressive Spatial Recurrent Neural Network (PS-RNN) \cite{ps-rnn} to progressively pass information along from preceding contents to the blocks to be encoded. 

Methods benefiting inter prediction were also proposed from many aspects. Yan \textit{et al.} proposed a Fractional Pixel Reference generation CNN (FRCNN) \cite{frcnn} to predict the fractional pixels inside the frame by adopting a three-layer CNN. Further, Liu \textit{et al.} proposed a Group Variation CNN (GVCNN) \cite{gvcnn} which can tackle multiple quantization parameters and sub-pixel positions in one model. Zhao \textit{et al.} proposed a method \cite{tcsct_inter} to enhance the inter-prediction quality by utilizing a CNN to combine two prediction blocks rather than a linear combination. Beyond PU-level combination, \cite{dvrf} and \cite{mascnn} directly exploited the learning capability of neural network to generate a new reference frame so that the residue of motion compensation can be greatly decreased.

Many efforts have also been made to in-loop filtering or post-processing. Park \textit{et al.} trained a shallow CNN for in-loop filtering firstly \cite{ivmsp_inloopfirst}. The network is inserted into HEVC codecs after DF with SAO off. Since then, many attempts have been made to enhance the representative capability of in-loop filtering networks. Dai \textit{et al.} proposed a Variable-Filter-Size Residual-Learning CNN (VRCNN) \cite{vrcnn} as the post-processing component with variable convolutional kernels to perceive multi-scale feature information. In \cite{icip18_mask}, He \textit{et al.} proposed a CNN adopting residual blocks for post-processing. In \cite{MIF}, Li \textit{et al.} proposed a Multi-frame In-loop Filter Network (MIF-Net) based on Dense Block \cite{densenet}. Wang \textit{et al.} proposed a Dense Residual CNN (DRN) \cite{drn} taking advantage of both dense shortcuts and residual learning. Also, many methods take intra-frame or inter-frame side information into consideration. In \cite{icip18_mask}, not only the decoded frames are sent into the network but also correspondent block partition side information. In \cite{strresnet}, Jia \textit{et al.} proposed a Spatial-Temporal Residue Network (STResNet) which aggregates temporal information by concatenating the feature maps of the co-located block and the current block together. In \cite{MIF}, a delicate reference frame selector was designed and the reference frames are warped by motion vectors predicted by neural network.

\subsection{Deep Learning Based Video Restoration}
With the surge of deep learning, video restoration also ushers in an outbreak. Many methods were first proposed to tackle image restoration such as denoising \cite{dncnn,ffdnet,denoise_cvpr19}, deraining \cite{derain_cvpr,derain_pami}, low-light enhancement \cite{low-light_1,low-light_2}, super-resolution \cite{srcnn,vdsr,srgan,edsr,rdn}, deblocking \cite{arcnn,cas-cnn,dmcnn} and so on. And these methods can be treated as single-frame restoration algorithms which don't utilize temporal redundancy of videos. To better fit the video scenario, many methods are proposed to utilize temporal information to help video restoration.

Incipient deep learning based video restoration works simply fuse frames together or concatenate feature maps together without motion compensation  \cite{vsr_tci16,strresnet}. 

Most common way to utilize temporal redundancy now is to warp reference frames to the current one by optical flow \cite{toflow,vsr_gcpr17,frvsr,MIF}. After that, the aligned frames will be send to neural networks to further reconstruct the current frame. While in \cite{rbpn}, Haris \textit{et al.} proposed a framework based on back-projection algorithm. Rather than aligning frames by flow, \cite{rbpn} directly sends the flow along with the reference frame together into the network without explicit alignment.

Another popular way to process temporal information is to pass hidden states frame by frame through a RNN module like LSTM \cite{lstm} or ConvLSTM \cite{convlstm}. In \cite{vsr_iccv17}, Tao \textit{et al.} proposed a sub-pixel motion compensation layer to provide finer motion compensation. Further, they proposed a ConvLSTM layer inside their network to pass temporal information. Beyond that, many variants from classic structures are proposed. In \cite{cvpr19_deblur}, the RNN cell is an Auto-Encoder structure which consists of multiple residual blocks. The hidden state is represented by the transformed feature maps extracted from the bottleneck in each RNN cell. 

Besides the mentioned methods, there exist other ways to handle temporal information. For example, in \cite{duf}, Jo \textit{et al.} utilized multi-frames to generate a dynamic upsampling filters to upsample low resolution frames. Lu \textit{et al.} used deep modules to substitute the original ones in Kalman Filter \cite{kalman} to process video sequences.

\section{Progressive Rethinking Networks \\ for In-Loop Filter}
\label{sec3}

In this section,
we at first present the motivation and design methodology of our proposed Progressive Rethinking Networks, \textit{i.e.} PR-CNN and PR-RNN.
Then, we discuss their detailed architectures step by step.

\subsection{Motivations}
\label{sec_Motivation}
In this paper, we aim to address the three issues of deep learning-based in-loop filters: 
1) Effective network design for feature learning;
2) Side information extraction and injection;
3) Joint spatial and temporal modeling in the coding context.
Our motivations to address these issues are three-fold:
\begin{itemize}
    \item \textbf{Representative Feature Refinement via Progressive Review.} 
    The basic blocks in previous advanced networks,
    \textit{e.g.} residue dense network (RDN)~\cite{rdn}, perform the progressive feature refinement. 
    However, at the end of each basic block, the feature dimension is compressed to avoid excessive growth of the model parameters, 
    which at the same time inevitably brings about the information loss across blocks.
    However, this information is also important.
    Intuitively, the high-dimensional feature is more informative (a record of total past experiences).
    After the compression, only most critical information (knowledge and principle) is preserved.
    When learning from new information, it will be helpful if the total past experiences are available.
    To this end, we introduce an inter-block connection that bypasses more information across blocks, which enables the model to learn by reviewing a compact representation of the complete past memorized experiences, namely ``rethinking''.
    \item \textbf{Hierarchical side information in the Coding Context.} 
    The coding process is performed block by block
	as the coding tree unfolds. Thus, the guidance side information should contain the partition-related side information to better represent the coding context, and guide the network to perform restoration from coarse to fine. 
	In this work, we extract MM-CU as side information to boost the proposed network for better in-loop filtering.
	
	\item \textbf{Collaborative Learning Mechanism.} Previous works make use of the temporal redundancy at the frame level (the aligned reference frame) unidirectionally.
	That is, they only make the information flow from the reference frames to the current one to improve its quality  without updating the state and feature of the reference frames. 
	In this paper, we propose a collaborative learning mechanism and maintain three learning paths to absorb useful information from reference frames (the nearest adjacent frame and peak quality frame) progressively and collaboratively.
	This design benefits acquiring useful information from the three kinds of resource and leads to a better restoration of the current reconstructed frame.
\end{itemize}

\subsection{Methodology Overview}
In modern video codecs, frames can be roughly divided into two categories according to whether temporal information is used. Similarly, our PRNs also include two versions, \textit{i.e.}  PR-CNN and PR-RNN, to process two kinds of reconstructed frames.
We will first introduce the pipeline of our method and then model these two versions step by step to develop the model more clearly.

\vspace{1mm}
\noindent {\textit{1) Pipeline}}

\vspace{1mm}
We first classify all frames into two categories, high-quality frame (\textit{H-frame}) and low-quality frame (\textit{L-frame}):
\begin{itemize}
    \item \textit{H-frame.} These frames include all I-frames,
    and each P-frame or B-frame whose POCs are multiples of 4.
    Based on the configuration of the codecs, these frames are usually compressed by lower quantization parameters (QP) and own higher quality.
    \item \textit{L-frame.} Other frames that do not belong to the first category, \textit{i.e.} P-frames and B-frames whose POCs are not multiples of 4, fall into L-frames as they are usually coded with fewer bits than H-frames.
\end{itemize}

\begin{figure}[tb]
	\centering
	\includegraphics[width=\linewidth]{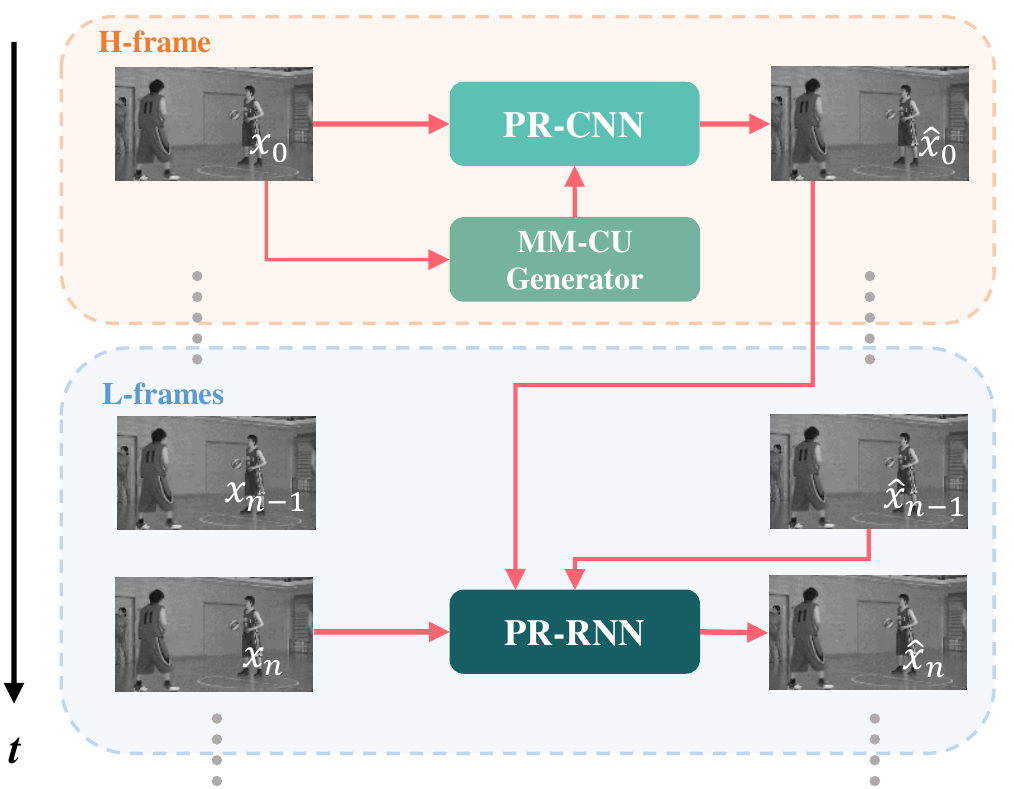}
	\caption{The pipeline of PRN. Under LD configuration, we filter the H frame, \textit{i.e} $x_0$, by PR-CNN with the guidance of its MM-CU side information maps. For the following frames, the filtered $\hat{x}_0$ is taken as a reference frame while the other reference frame is the nearest restored frame in temporal domain which is simply the previous frame under LD configuration. Therefore, for $x_0$, it selects $\hat{x}_{n-1}$ and $\hat{x}_0$ as its reference frames.}
	
	\label{fig:Overview}
\end{figure}

Our network uses PR-CNN and PR-RNN to filter H-frames and L-frames, respectively.
Our pipeline under LD configuration is shown in Fig.~\ref{fig:Overview}. 
The reason to process two kinds of frames differently is that, for H-frames, reference frames often have lower quality and may consequently mislead its restoration. Therefore, we only take intra-frame information, \textit{i.e.} MM-CU, as side information to help filtering. 
In Fig. \ref{fig:Overview}, PR-CNN takes $x_0$ and its MM-CU maps as input and outputs the filtered result $\hat{x}_0$.
After that, $\hat{x}_0$ is taken as the reference frame of the successive L-frames. 
Besides $\hat{x}_0$, for each L-frame, PR-RNN also takes the filtered neighboring frame $\hat{x}_{n-1}$ as another input because the neighboring frame contains most shared content information. 
Therefore, we in all take 3 frames as the input of PR-RNN, \textit{i.e.} the current frame, the neighboring frame and an H-frame.
When $n-1$ is the multiple of 4, \textit{i.e.}  $\hat{x}_{n-1}$ is also an H-frame, we simply take this frame as the neighboring frame without exceptional operations.
Under the RA configuration, the pipeline is quite similar except that the coding order of frames is different. To be specific, H-frames are still filtered by PR-CNN and L-frames are filtered by PR-RNN. The neighboring reference frame is not exactly the previous frame in temporal domain. We select the most neighboring frame from the decoded frames buffer as the neighboring reference frame. Under the AI configuration, all frames are filtered by PR-CNN as no inter-frame correlation is guaranteed.


\vspace{1mm}
\noindent {\textit{2) Modeling PRN Step by Step}}
\vspace{1mm}

To provide a better understanding on our model design,  we construct our deep network step by step.

\begin{itemize}
\item\textbf{Residual Dense Network}. 
We take a previous excellent work  residual dense network (RDN)~\cite{rdn} as the starting point of our model.
As shown in Fig.~\ref{fig:Steps}(a),
a series of residual dense blocks (RDB) are stacked.
There is an additional bypass connection to link the first and later layers to better trade-off between the local and global signal modeling.

\item\textbf{PR-CNN}. As shown in Fig.~\ref{fig:Steps}(b), different from RDN~\cite{rdn}, inter-block connections (red line) are added to bypass richer information across blocks.
These connections are non-trivial, as they make the successive blocks ``rethink'', namely, learning to extract more representative features guided by the previous information without dimension compression.
Furthermore, we inject the side information into the network to facilitate in-loop filter.
MM-CU maps are extracted and used as another input to guide the restoration process.

\begin{figure}
    \centering
    \includegraphics[width=\linewidth]{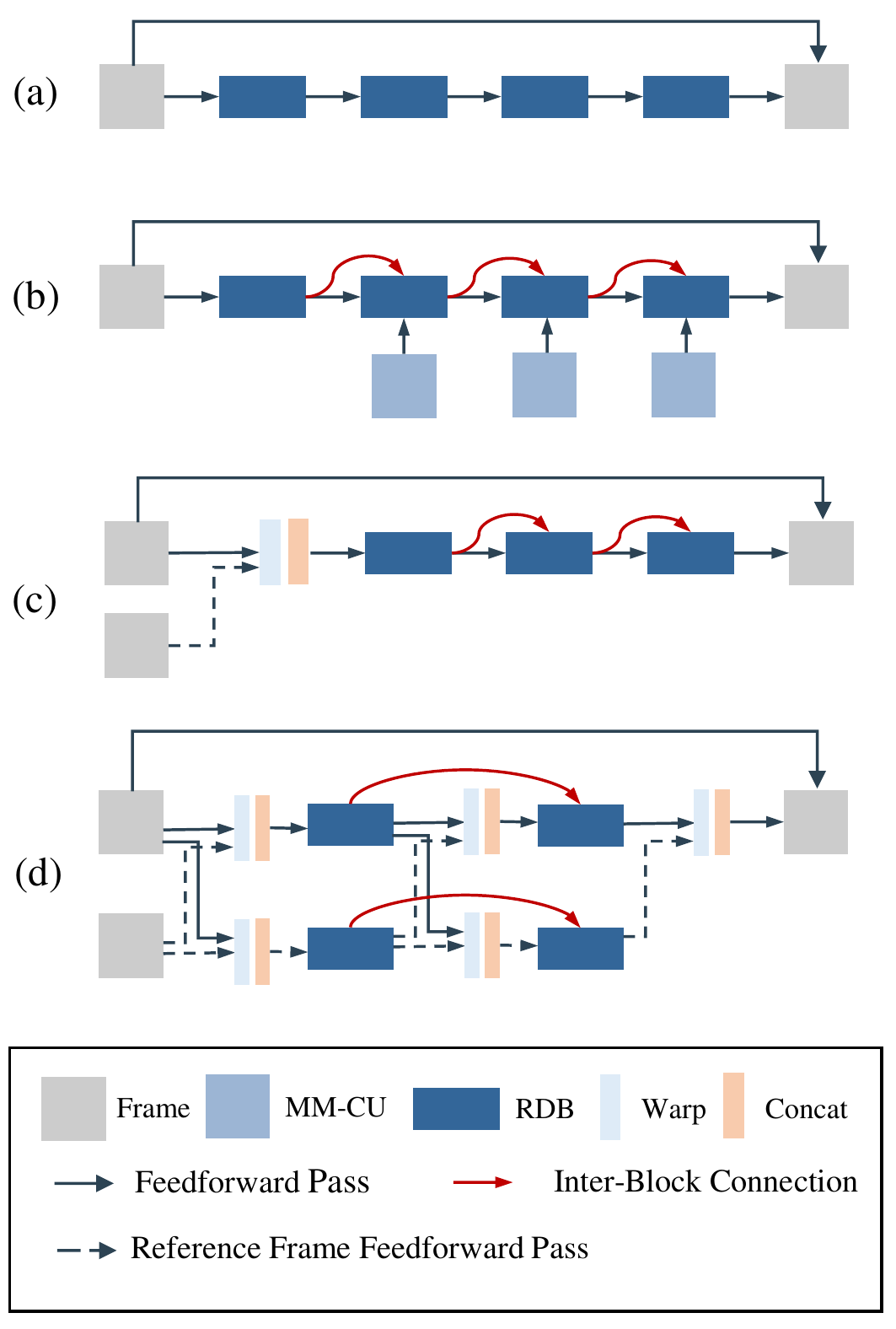}
    \caption{Modeling PR-CNN and PR-RNN step by step. (a) The simplified architecture of RDN \cite{rdn} which consists of RDBs. (b) By adding inter-block connection and side information maps, we establish the PR-CNN. (c) The most common way to utilize temporal information: warping the reference frame to the current one and then concatenating them together as the input of the network. (d) The architecture of PR-RNN with only two states. Collaborative learning mechanism is built with feature sharing and synchronous state updating. }
    \label{fig:Steps}
\end{figure}

\item\textbf{Frame-Level Temporal Fusion.} 
To exploit the temporal redundancy of video frames, the commonly used way in previous methods is shown in Fig.~\ref{fig:Steps}(c).
The network takes the warped reference frame as the input.
However, this way might not make full use of the temporal dependencies.

\item\textbf{PR-RNN (Feature-Level Aggregation and Collaborative Learning).}
Beyond taking the aligned reference frame as the input, 
we further develop a collaborative learning mechanism to 
exploit temporal 
dependencies bidirectionally at the feature level.
Specifically, the feature map is also feed-forwarded from the reference frame to the current one as shown in Fig.~\ref{fig:Steps}(d).
It is noted that, in our implementation, we use recurrent neural modules to update the feature maps 
and Fig.~\ref{fig:Steps}(d) is a simplified unfolding version of our proposed PR-RNN. 
\end{itemize}

In the following, we will present our PRN in details, including its basic module PRB, and PR-CNN as well as PR-RNN.

\subsection{Progressive Rethinking Block}
\label{sec_PRB}

To fully utilize the past memory for current restoration, we design a Progressive Rethinking Block (PRB), which has an additional inter-block connection to forward more informative feature representations across blocks. 
The structure of our proposed PRB is shown in Fig.~\ref{fig:PRN}(b).

In an RDB, 
the input feature map $F_{d-1}$ is first feed-forwarded to a series of  convolutional layers to extract rich hierarchical features, and the ReLU activation layers are injected between convolutional layers to model nonlinearity.
The procedure is formulated as follows,
\begin{equation}
	G_d = H_d(F_{d-1}),
\end{equation}
where $H_d\left( \cdot \right)$ is the corresponding  nonlinear transform procedure.
The concatenated hierarchical feature (accumulated in the way of dense connections from $H_d$) is denoted as $G_d$.
As the channel dimension of $G_d$ is greatly larger than the input $F_{d-1}$, we compress the channel dimension via a 1$\times$1 convolutional layer, 
and the residual connection can be utilized to accelerate convergence. However, the dimension compression inevitably causes information loss.  

To compensate for this loss, 
we introduce another path to send the feature map of the previous block $M_{d-1}$ to that of the current block simply by concatenating it with $G_d$ denoted by red lines in Fig.~\ref{fig:PRN}(b).
This connection is nontrivial as with it, all modules, \textit{i.e.} PRBs, are connected with a feature path that keeps the high-dimensional informative features from bottom to top. Therefore, the generation of the features at one PRB is guided by both the compressed feature at the last PRB and the previous forwarded high-dimensional feature, which critically provides more abundant low-level features to facilitate more powerful feature learning in the current PRB. Besides, with the inter-block connection, all PRBs will have a higher dimensional feature representation space, where better features are easy to be obtained throughout a thorough training process.
We generate $M_d$ by a $1\times1$ convolutional layer as follows,
\begin{equation}
    M_{d} = P_M(\left[G_d, M_{d-1}\right]),
\end{equation}
where $P_M(\cdot)$ is the corresponding process,
and $[\cdot]$ denotes the concatenation operation. Similarly, we can generate $F_d$ and add a local residual learning for better gradient back-propagation as follows,
\begin{equation}
    F_d=P_F([G_d,M_d-1])+F_{d-1},
\end{equation}
where $P_F(\cdot)$ is also a $1\times1$ convolutional function.

As a summary, we can conclude the process of PRB as \text{$\text{P}_{\text{PRB}} \left( \cdot \right)$} and for the k-\textit{th} PRB, there exists
\begin{equation}
\label{f1}
	\left[F_k , M_k\right] = \text{P}_\text{PRB}(F_{k-1}, M_{k-1}).
\end{equation}

\subsection{Progressive Rethinking Convolutional Neural Network}

\begin{figure*}[tb]
	\centering
	\includegraphics[width=\linewidth]{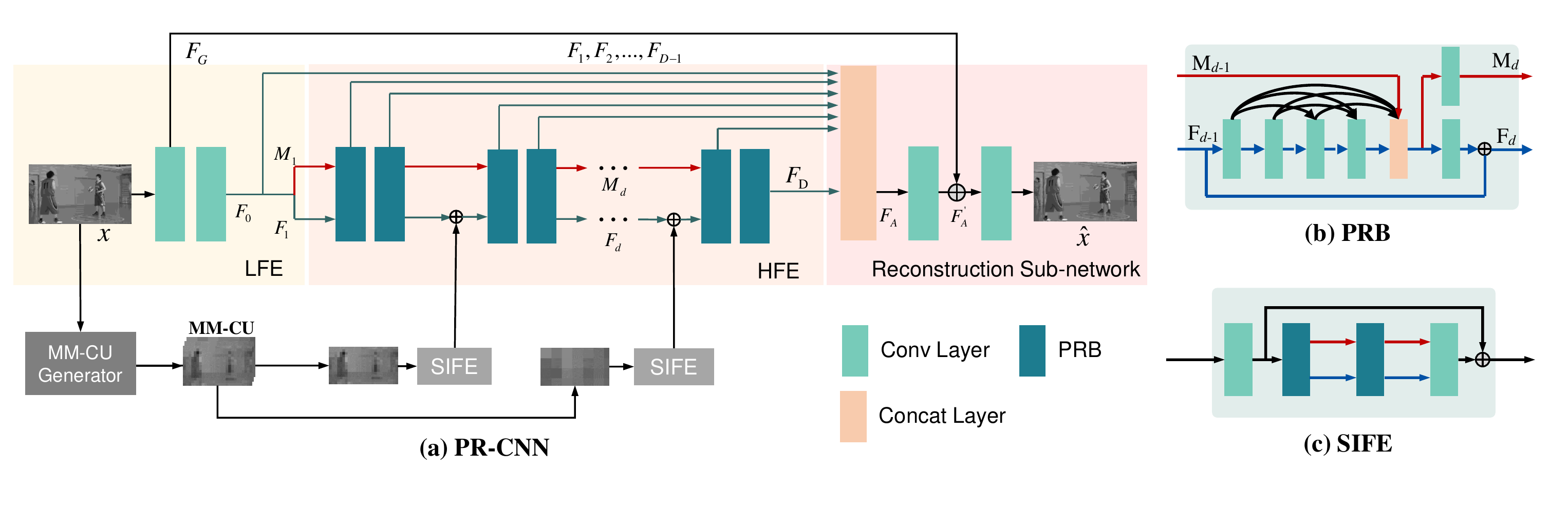}
	
	\caption{(a) The architecture of the Progressive Rethinking Convolutional Neural Network. The network takes the unfiltered frame as the input and it generates the filtered output frame. The feature maps extracted from the side information by SIFE are added to data flow during the processing. (b) The structure of Progressive Rethinking Block. (c) The architecture of Side Information Feature Extractor.}
	
	\label{fig:PRN}
\end{figure*}

To process H-frames, which are usually with high quality, 
we only make use of spatial redundancy and the related side information for in-loop filter of the corresponding reconstructed frames. The overall architecture of our PR-CNN is shown in Fig.~\ref{fig:PRN}(a). 
It has two branches: the main brunch, \textit{i.e.} the PR-CNN baseline network without MM-CU maps, and side information feature extractor (SIFE). We will illustrate their architectures in details.

\vspace{1mm}
\noindent {\textit{1) Architecture of Main Branch of PR-CNN}}
\vspace{1mm}

PR-CNN takes the unfiltered frame $x$ and MM-CU maps as its input.
$x$ is fed to the main brunch,
and MM-CU maps are first fed into SIFE and then fused to the main brunch. 
PR-CNN can be roughly divided into 3 parts: \textit{Low-Level Feature Extractor (LFE), High-Level Feature Extractor (HFE) with MM-CU Fusion and Reconstruction Sub-Network}.
\vspace{1mm}

\noindent\textbf{Low-Level Feature Extractor.} 
The input frame is first fed into a Low-level Feature Extractor for low-level feature extraction.
The LFE consists of two convolutional layers. The corresponding process is formulated  denoted as $\text{P}_{\text{LFE}}(\cdot)$:
\begin{equation}
    F_0 = \text{P}_\text{LFE}(x),
\end{equation}
where $F_0$ is the generated feature maps.
\vspace{1mm}

\noindent\textbf{High-Level Feature Extractor with MM-CU Fusion.}
$F_0$ is further feed-forwarded into $D$ sequential PRBs, namely High-level Feature Extractor. 
It is noted that, each PRB indeed needs two inputs: $M_k$ and $F_k$ as shown in Eqn.~\eqref{f1}. 
We initially set $M_0=F_0$. 
After a certain number of PRBs, we fuse the feature maps of a Mean value of CU (M-CU) into the main brunch by element-wise addition. We use $SF_k$ to denote feature map of the k-\textit{th} M-CU, and it is inserted to the main branch after the $n_k$-\textit{th} PRB. The process is denoted as follows,
\begin{equation}
    F_{n_{k}}=F_{n_{k}}+SF_{k}.
\end{equation}
\vspace{1mm}

\noindent\textbf{Reconstruction Sub-network.} After $D$ PRBs, we concatenate all feature maps $\{F_1,F_2,..., F_D\}$ together and use a 1$\times$1 convolutional layer denoted as $\text{P}_{\text{Compress}}(\cdot)$ to compress them as follows:
\begin{equation}
	F_{C} = \text{P}_\text{Compress}\left(\left[F_1,F_2,F_3...,F_D\right]\right).
\end{equation}

We then append a global residual connection from the first convolutional layer $F_G$ to the last one as follows,
\begin{equation}
    F_{C}^{'} = F_{C}+F_G.
\end{equation}

Finally, we construct the output $\hat{x}$ by a 3$\times$3 convolutional layer denoted as $\text{P}_{\text{Rec}}(\cdot)$:
\begin{equation}
    \hat{x} = \text{P}_{\text{Rec}}(F_{C}^{'}).
\end{equation}

\begin{figure}[tb]
		\centering
		\subfigure[Partition Tree]{
			\begin{overpic}[width=80mm]{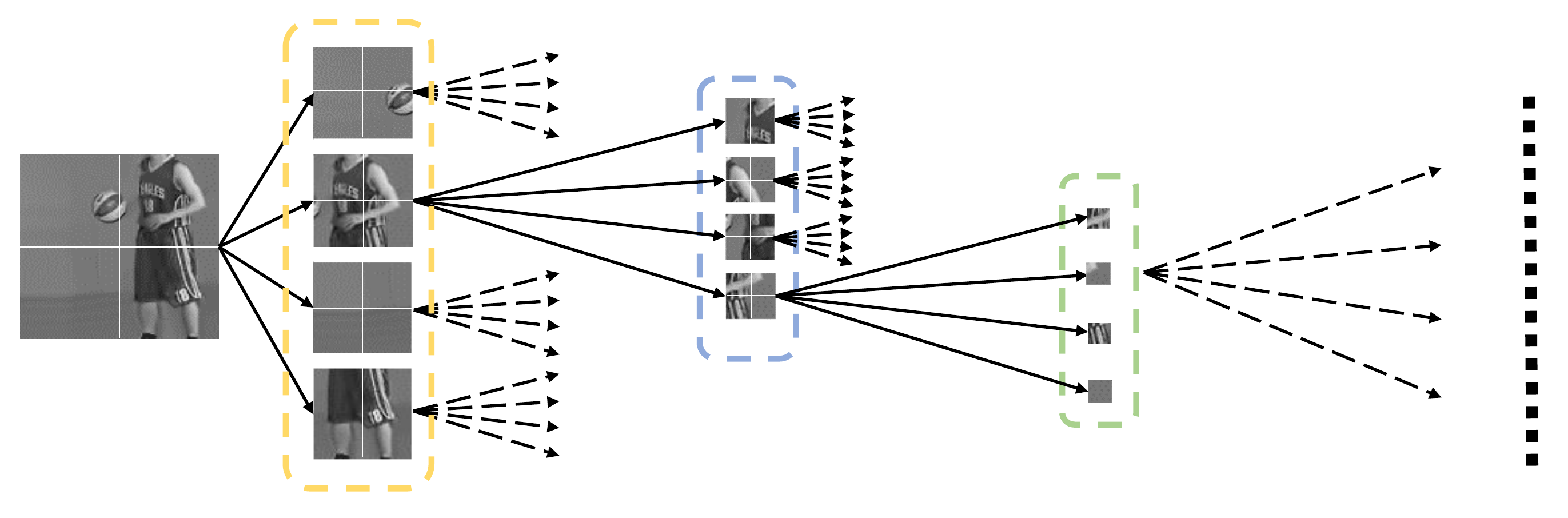}\end{overpic}}
		\subfigure[Multi-Scale Mean value of CU]{
			\begin{overpic}[width=80mm]{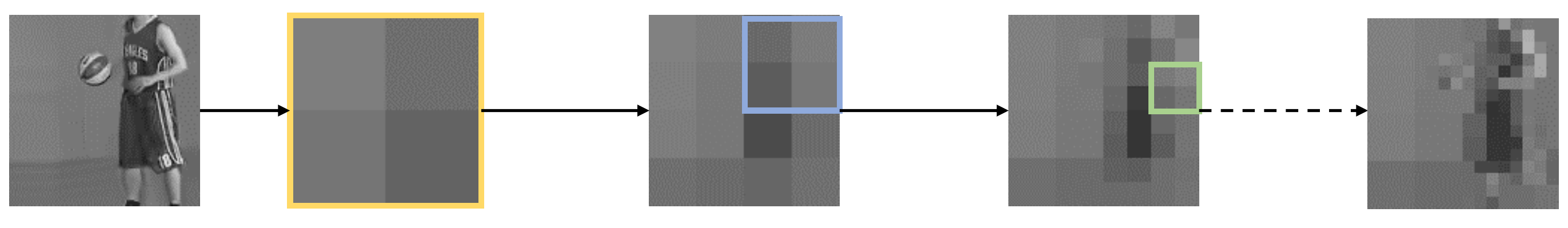}\end{overpic}}
			
		\caption{(a) A partition tree generated by the codec. We calculate the mean value of each CU at different levels as the side information maps. (b) The visualized results of side information maps of different scales.}
		
		\label{fig:Partition Tree}
\end{figure}

\vspace{1mm}
\noindent \textit{2) MM-CU Generation and Fusion}
\vspace{1mm}

In addition to only utilizing the frame information, we further fuse intra-frame side information extracted from the HEVC codec into our network. 
As HEVC encodes a frame at the CU level independently with different coding parameters, the partition information contains a lot of extra important side information which is beneficial for in-loop filter.
\vspace{1mm}

\noindent\textbf{MM-CU Generation.} 
Different from only generating M-CU at the bottom layer (leaf node of the partition tree) of the quadtree~\cite{icip18_mask}, we also extract M-CU in the intermediate layers (every node of the partition tree).
Namely, we calculate the mean value of a CU every time a partition happens. 
Consequently, the side information includes the information related to the entire coding partition architecture, and therefore guides the network to remove the coding artifacts from coarse to fine.

We calculate the mean value of each CU at different levels from coarse to fine to derive the corresponding side information maps. 
As shown in Fig.~\ref{fig:Partition Tree}(b), blocks in the yellow dotted box are four CTUs and their corresponding M-CU side information maps, and the coarsest ones are surrounded by a yellow border.
Then, every time the CUs are divided into four smaller CUs, we recursively calculate the mean value of each partitioned CU. If the CU is not divided, we keep its side information value the same as that at the upper level, namely that the side information value of that CU is unchanged.
The recursive process stops when the CU cannot be partitioned anymore.
Finally, the multi-scale M-CU side information maps, MM-CUs, are obtained.
\vspace{1mm}

\noindent\textbf{MM-CU Fusion.}
The information of MM-CU is first transformed into the feature map, and then injected into different layers of the main branch.
The feature of each M-CU is extracted by a simple shallow CNN named side information Feature Extractor (SIFE), whose structure is shown in Fig.~\ref{fig:PRN}(c).
The M-CU first goes through a convolutional layer and two stacked PRBs. 
After that, a residual connection is added. At last, a convolutional layer generates the final output feature map of the M-CU. It is intuitive that, the information of finer M-CU maps reflects local details of the coding architecture more 
while that of coarser ones contain more global coding structure information. 
Thus, we inject coarser M-CU maps to the main branch 
in deeper layers 
so that the global information can play a more important role in guiding the network training when larger areas are perceived in deeper layers.
We choose the element-wise addition as the fusion operation.

\begin{figure*}[tb]
	\centering
	\includegraphics[width=\linewidth]{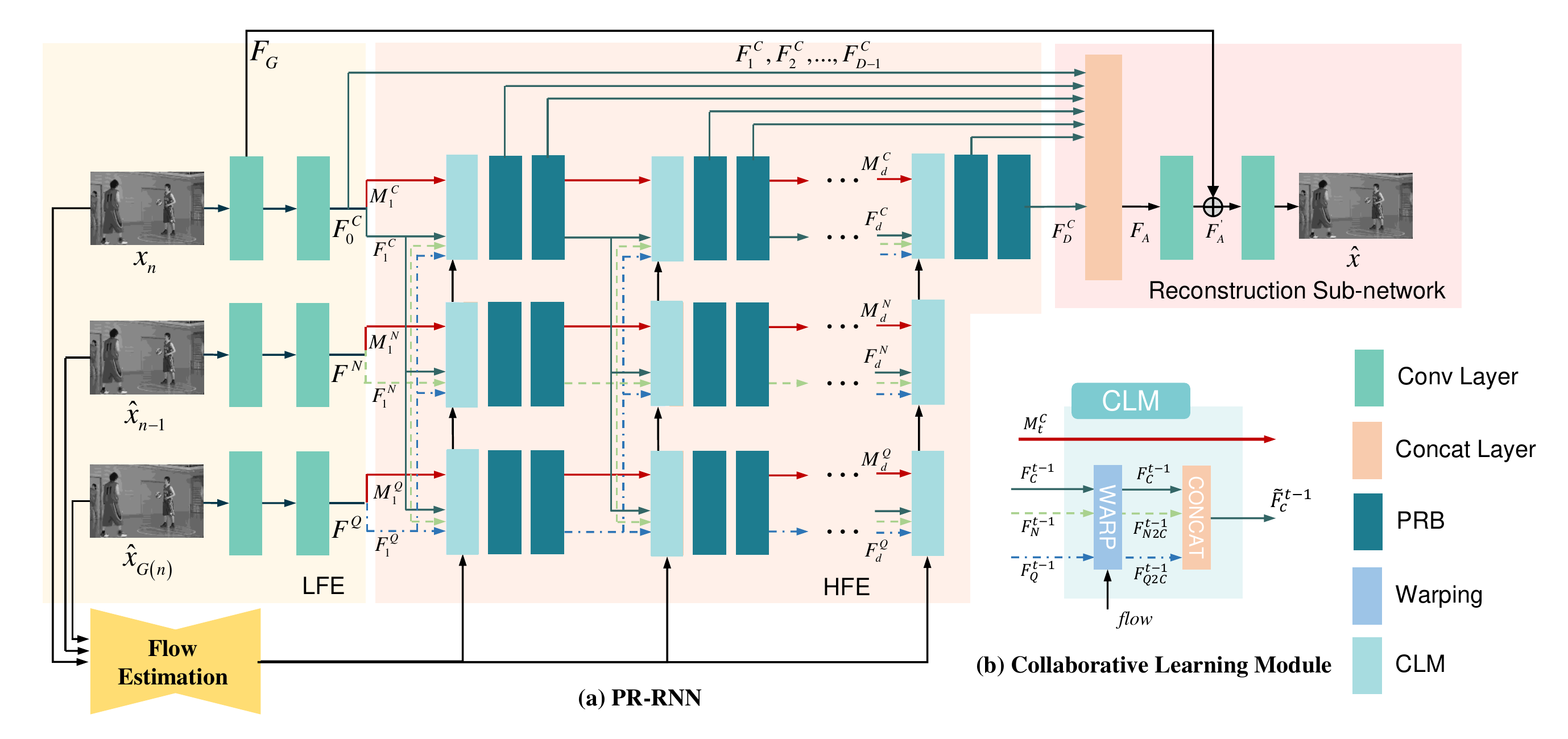}
	
	\caption{(a) The architecture of the Progressive Rethinking Recurrent Neural Network under LD configuration. (b) The structure of Collaborative Learning Module. }
	
	\label{fig:PR-RNN}
\end{figure*}

\subsection{Progressive Rethinking Recurrent Neural Network}

Besides exploiting the current frame information, we further develop a Progressive Rethinking Recurrent Neural Network (PR-RNN) to effectively utilize inter-frame side information with a collaborative learning. 
We will first provide the network architecture of PR-RNN and then introduce the collaborative learning in detail.

\vspace{1mm}
\noindent \textit{1) Architecture of PR-RNN}
\vspace{1mm}

The architecture of our PR-RNN is shown in Fig.~\ref{fig:PR-RNN}(a). To clearly show the relationship of PR-RNN with PR-CNN, we show an unfolding version of PR-RNN. Different from PR-CNN, PR-RNN generates a filtered frame with the information of both  current frame and reference frames (nearest adjacent frame and peak quality frame) to further improve frame quality when inter-prediction is available. Specifically, PR-RNN takes three kinds of frames as its input:
\begin{itemize}
    \item \textit{Current Frame $x_n$.}
    \item \textit{Neighboring Frame.} We select the nearest filtered frame in the temporal domain as another input of the network as it is often the most similar frame in all reconstructed frames to the current frame. 
    Under the LD configuration, 
    it will be the last filtered frame $\hat{x}_{n-1}$ as shown in Fig.~\ref{fig:PR-RNN}. Under RA configuration, it is a little more complex because the frames are not coded in a sequential order. We just still choose the nearest filtered frame as one of the reference frame.
    \item \textit{Peak Quality Frame.} We also choose the nearest filtered H-frame as another reference frame. 
    More high-frequency information is preserved in this frame,  which benefits the restoration of the current frame. 
    It is denoted as $\hat{x}_{G(n)}$ in Fig.~\ref{fig:PR-RNN} and $G(n)$ is denoted as follows,
    \begin{equation}
        G(n) = n-(n~mod~GOP\_SIZE).
    \end{equation}
\end{itemize}

To apply the in-loop filtering frame-by-frame along the temporal dimension, the three input frames at different temporal steps make up three queues, which we abstract into three states: \textit{State C, State N} and \textit{State Q} to denote the  \textbf{\textit{C}}urrent frame, \textbf{\textit{N}}eighbouring frame, peak \textbf{\textit{Q}}uality frame respectively. Therefore, we can also use $x_C$, $x_N$ and $x_Q$ to represent the three frames in these state queues, respectively.

PR-RNN can be divided into four parts: 
\textit{Flow Estimation}, 
\textit{Low-Level Feature Extractor},
\textit{Recurrent Module with Collaborative Learning Mechanism}, 
and  \textit{Reconstruction Sub-Network}.
\vspace{1mm}

\noindent\textbf{Flow Estimation.} Because the three frames are not aligned, we estimate their optical flow results and apply warping operations. 
We adopt SpyNet \cite{spynet} to generate the optical flow maps. We use $i$ and $j$ to represent any two states and we can get:
\begin{equation}
    \text{flow}_{i \to j} = \text{P}_\text{SpyNet}(x_j, x_i),
\end{equation}
where the first parameter of the function \text{$\text{P}_{\text{SpyNet}} \left( \cdot \right)$} is the target frame and the second one is the source frame.
\vspace{1mm}

\noindent \textbf{Low-level Feature Extractor.} Instead of just warping all frames to the current frame, we need to extract the low-level feature maps of the three inputs and further warping them in our recurrent module by a collaborative learning mechanism. The extraction of low-level feature is same as the one in PR-CNN.
However, we name the corresponding process as  $\text{P}_{\text{RLFE}}(\cdot)$ to highlight that it belongs to PR-RNN. Therefore, we can get:
\begin{equation}
    F_i^0 = \text{P}_{\text{RLFE}}(x_i),
\end{equation}
where $F_i^j$ denotes the feature map of $\textit{\text{State}} ~i\in\{\textit{\text{State}}~C,~\textit{\text{State}}~N,~ \textit{\text{State}}~Q\}$ after $j$ times unfolding. $x_i$ is the frame that corresponds to $\textit{\text{State}}~i$.
\vspace{1mm}

\noindent \textbf{Recurrent Module with Collaborative Learning.} After the flow estimation and low-level feature extraction, the feature maps of these three states and their flow maps are fed into the recurrent module for collaborative learning. Namely, the input feature maps are processed by the Collaborative Learning Module (CLM) and then pass sequential PRBs to further update the state. The detailed process of this collaborative learning will be introduced in the next subsection.
It should be mentioned that the feature map of $\textit{\text{State}}~C$ is temporarily kept at each time-step and they are concatenated together and fed into the successive layers as follows,
\begin{equation}
    F_{C} = \left [F_C^0, F_C^1, ..., F_C^T\right],
\end{equation}
where $T$ is the total unfolding times of our PR-RNN.
\vspace{1mm}

\noindent\textbf{Reconstruction Sub-network.} At last, we reconstruct the frame through two convolutional layers. The first convolutional layer is $1\times1$ to compress the channel number. Then, a global residual connection is used to connect the first convolutional layer $F_G$ and the last layer as follows,
\begin{equation}
    F_{C}^{'} = F_{C} + F_G.
\end{equation}

The final output result is reconstructed by a $3\times3$ convolutional layer. The process of the overall reconstruction sub-network can be denoted as follows,
\begin{equation}
    \hat{x}_C = \text{P}_{\text{RRec}}(F_{C}^{'}), 
\end{equation}
where $\text{P}_{\text{RRec}}(\cdot)$ stands for the convolutional function. 


\begin{table*}[htb]
    \centering
    \caption{Overall experimental results on class A to class E. Only BD-rate of Y channel is shown.}
\begin{tabular}{c|l|c|c|c|c}
\hline
\textbf{Class} & \multicolumn{1}{c|}{\textbf{Sequence}} & \textbf{All-Intra} & \textbf{Low-Delay B} & \textbf{Low-Delay P} & \textbf{Random-Access} \bigstrut\\
\hline
\hline
\multirow{5}[4]{*}{\textbf{A}} & SteamLocomotiveTrain & -1.3\% & -     & -     & -5.4\% \bigstrut[t]\\
      & Traffic & -10.9\% & -     & -     & -8.7\% \\
      & Netuba & -3.2\% & -     & -     & -1.7\% \\
      & PeopleOnStreet & -9.7\% & -     & -     & -9.5\% \bigstrut[b]\\
\cline{2-6}      & Average & -6.3\% & -     & -     & -6.3\% \bigstrut\\
\hline
\hline
\multirow{6}[4]{*}{\textbf{B}} & Kimono & -7.4\% & -5.8\% & -8.9\% & -5.0\% \bigstrut[t]\\
      & ParkScene & -7.9\% & -10.5\% & -5.4\% & -5.0\% \\
      & BasketballDrive & -7.9\% & -4.6\% & -12.4\% & -10.3\% \\
      & BQTerrace & -2.5\% & -8.4\% & -15.0\% & -10.4\% \\
      & Cactus & -7.5\% & -9.6\% & -10.4\% & -8.7\% \bigstrut[b]\\
\cline{2-6}      & Average & -6.6\% & -7.8\% & -10.4\% & -7.9\% \bigstrut\\
\hline
\hline
\multirow{5}[4]{*}{\textbf{C}} & BasketballDrill & -17.4\% & -10.9\% & -12.3\% & -9.9\% \bigstrut[t]\\
      & BQMall & -11.2\% & -10.2\% & -10.9\% & -9.0\% \\
      & PartyScene & -6.8\% & -6.0\% & -7.5\% & -5.7\% \\
      & RaceHorsesC & -7.3\% & -9.0\% & -9.4\% & -8.2\% \bigstrut[b]\\
\cline{2-6}      & Average & -10.7\% & -9.0\% & -10.0\% & -8.2\% \bigstrut\\
\hline
\hline
\multirow{5}[4]{*}{\textbf{D}} & BasketballPass & -11.1\% & -8.6\% & -9.0\% & -7.8\% \bigstrut[t]\\
      & BlowingBubbles & -7.7\% & -4.9\% & -5.7\% & -4.3\% \\
      & BQSquare & -8.6\% & -6.9\% & -9.1\% & -7.6\% \\
      & RaceHorses & -11.0\% & -9.8\% & -9.7\% & -8.5\% \bigstrut[b]\\
\cline{2-6}      & Average & -9.6\% & -7.6\% & -8.4\% & -7.1\% \bigstrut\\
\hline
\hline
\multirow{4}[4]{*}{\textbf{E}} & FourPeople & -14.2\% & -12.7\% & -12.8\% & -11.4\% \bigstrut[t]\\
      & Johnny & -13.4\% & -13.1\% & -17.6\% & -12.8\% \\
      & KristenAndSara & -12.4\% & -13.3\% & -14.3\% & -10.4\% \bigstrut[b]\\
\cline{2-6}      & Average & -13.3\% & -13.0\% & -14.9\% & -11.5\% \bigstrut\\
\hline
\hline
\textbf{ALL} & \textbf{Average} & \textbf{-9.0\%} & \textbf{-9.0\%} & \textbf{-10.6\%} & \textbf{-8.0\%} \bigstrut\\
\hline
\end{tabular}
    \label{tab:overall}
\end{table*}

\vspace{1mm}
\noindent \textit{2) Collaborative Learning}
\vspace{1mm}

We will illustrate the collaborative learning in detail. We apply the collaborative learning mechanism through a Collaborative Learning Module as Fig. \ref{fig:PR-RNN}(a) shows. The detailed structure of CLM is shown in Fig. \ref{fig:PR-RNN}(b).
At time-step $t$, the feature maps that correspond to the three states are updated as follows,
\begin{equation}
    [F_C^{t}, F_N^{t}, F_Q^{t}] = \text{P}_{\text{RM}}(F_C^{t-1}, F_N^{t-1}, F_Q^{t-1},\text{flow}),
\end{equation}
where $F_C^t$ denotes the feature maps of $\textit{\text{State}}~C$ at time-step $t$. Similarly, $F_N^t$ and $F_Q^t$ stand for the feature maps of $\textit{\text{State}}~N$ and $\textit{\text{State}}~Q$ at time-step $t$, respectively. $ \text{P}_{\text{RM}}\left( \cdot \right)$ is the mapping of our recurrent module.

To be specific, the three states are first warped with the estimated flow maps as follows,
\begin{equation}
    F_{i2j}^{t-1} = \text{warp}(F_i^{t-1}, \text{flow}_{i \to j}),
\end{equation}
where $i$ and $j$ represent two arbitrary states, respectively.

Then, the warped feature maps are concatenated to interact and share information with each other as follows,
\begin{align}
    {\widetilde{F}_C}^{t-1} = [F_C^{t-1}, F_{N2C}^{t-1},F_{Q2C}^{t-1}], \nonumber
	\\ 
    {\widetilde{F}_N}^{t-1} = [F_{C2N}^{t-1}, F_N^{t-1},F_{Q2N}^{t-1}],
	\\
    {\widetilde{F}_Q}^{t-1} = [F_{C2Q}^{t-1}, F_{N2Q}^{t-1},F_Q^{t-1}]. \nonumber
\end{align}


After collaborative information sharing, the features of the three states are first compressed by a $1\times1$ convolutional layer and further refined by several PRBs. We denote the corresponding process as $\text{P}_{\text{Tr}}\left( \cdot \right)$. Therefore, the procedure can be formulated as follows,
\begin{equation}
    F_i^t = \text{P}_{\text{Tr}}({\widetilde{F}_i}^{t-1}).
\end{equation}

Till now, three states are all updated. 
$F_i^t$ is further fed into the next recurrence to improve the restoration quality progressively. 

\section{Implementation Details}
\label{sec4}
\subsection{Network Implementation}
The PR-CNN is made up of 10 PRBs. As our anchor HEVC codec is HM 16.15, which only provides us a 4 layer partition tree, our MM-CU maps consist of 4 different scale M-CUs. We insert the feature maps of the M-CUs after 2-\textit{nd}, 4-\textit{th}, 6-\textit{th}, 8-\textit{th} PRB respectively from fine to coarse.

The PR-RNN has three states as we have mentioned above. For each state, the respective recurrent module is made up of 3 PRBs. Therefore, there are 9 PRBs in PR-RNN in all. The folding time $T$ is set to 2.

All activation functions in our PRNs are ReLU. The kernel of each convolution layer is 3$\times$3 except that the kernel of the emphasized channel compression module after concatenation layer is $1\times1$.

\subsection{Training} 
We train PR-CNN and PR-RNN on DIV2K \cite{div2k} and Vimeo-90K \cite{toflow}, respectively. The DIV2K dataset contains 800 diverse high-resolution images while Vimeo-90K contains 89,800 clips with 7 frames. We randomly extract 18,345 clips from Vimeo-90K with 4 frames in each clip.

We crop the image into 64$\times$64 and 128$\times$128 patches for the training of PR-CNN and PR-RNN. We apply random flipping both vertically and horizontally for augmentation. 

The network is implemented in Pytorch and Adam is used as the optimizer with $\beta_1 = 0.9, \beta_2 = 0.999$. The learning rate is first set to 0.0001 and adaptively decreased until convergence. We train one model for each QP. We first train PR-CNN and PR-RNN suffering the worst degradation (QP 37) for 75 and 40 epochs respectively and then finetune other models from them for 20 epochs.

\subsection{Integration}
We insert our PR-CNN and PR-RNN between DF and SAO modules. Only luma component is filtered by our method.

We also adopt CTU level RDO under LDP, LDB and RA configurations to choose whether to use filtered results or unfiltered results. While under AI configuration, we simply substitute our filtered frame for the unfiltered frame without RDO.

\section{Experimental Results}
\label{sec5}
In this section, we will show the experimental results of our models. As mentioned in the previous section, we utilize PR-CNN to  filter \textit{H-frames} and PR-RNN to filter \textit{L-frames}. The testing QPs belong to \{22,27,32,37\}.

\subsection{Overall performance}
\begin{figure*}[tbp]
	\centering
\subfigure[All Intra]{
	\includegraphics[width=0.23\linewidth]{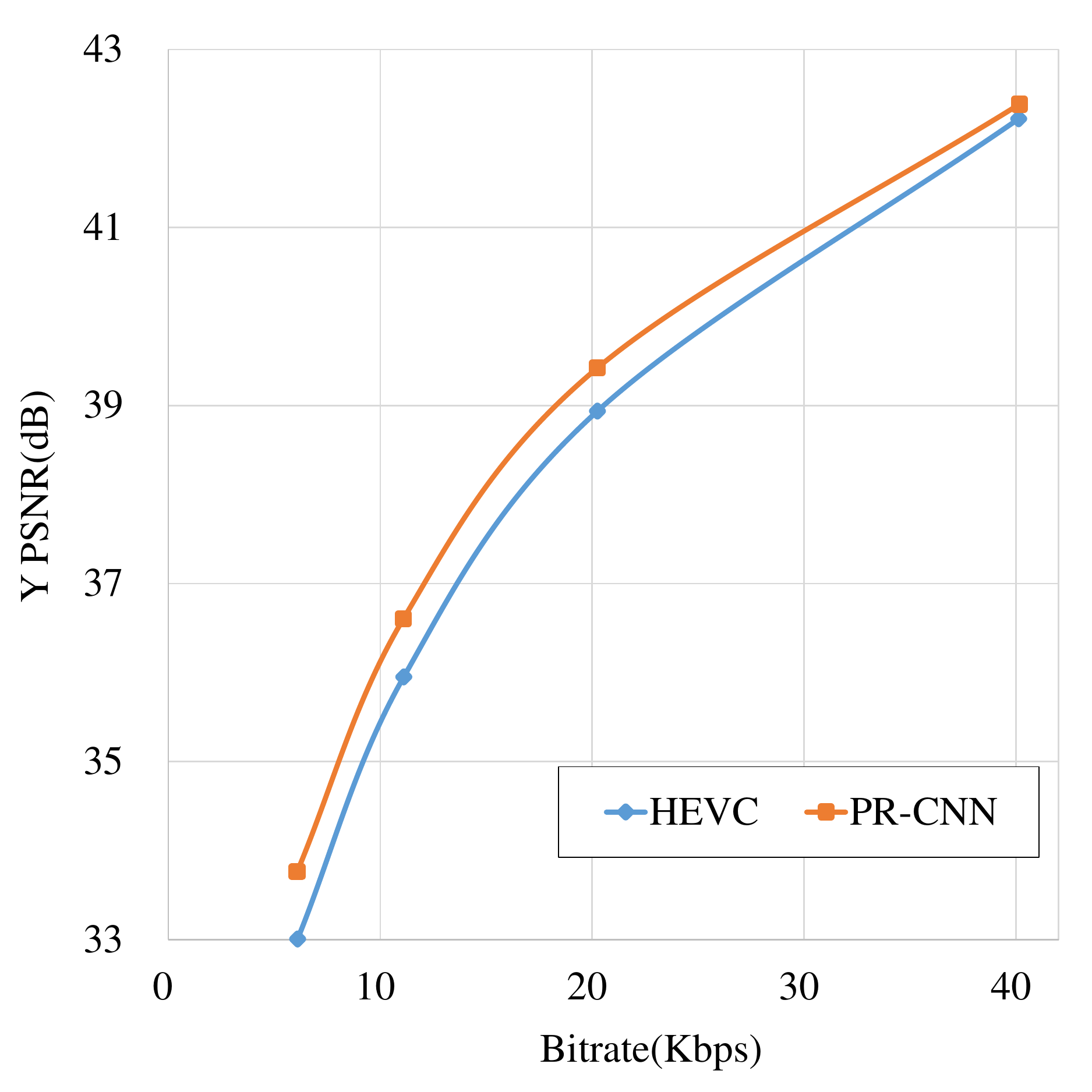}}
\subfigure[Low-Delay B]{
	\includegraphics[width=0.23\linewidth]{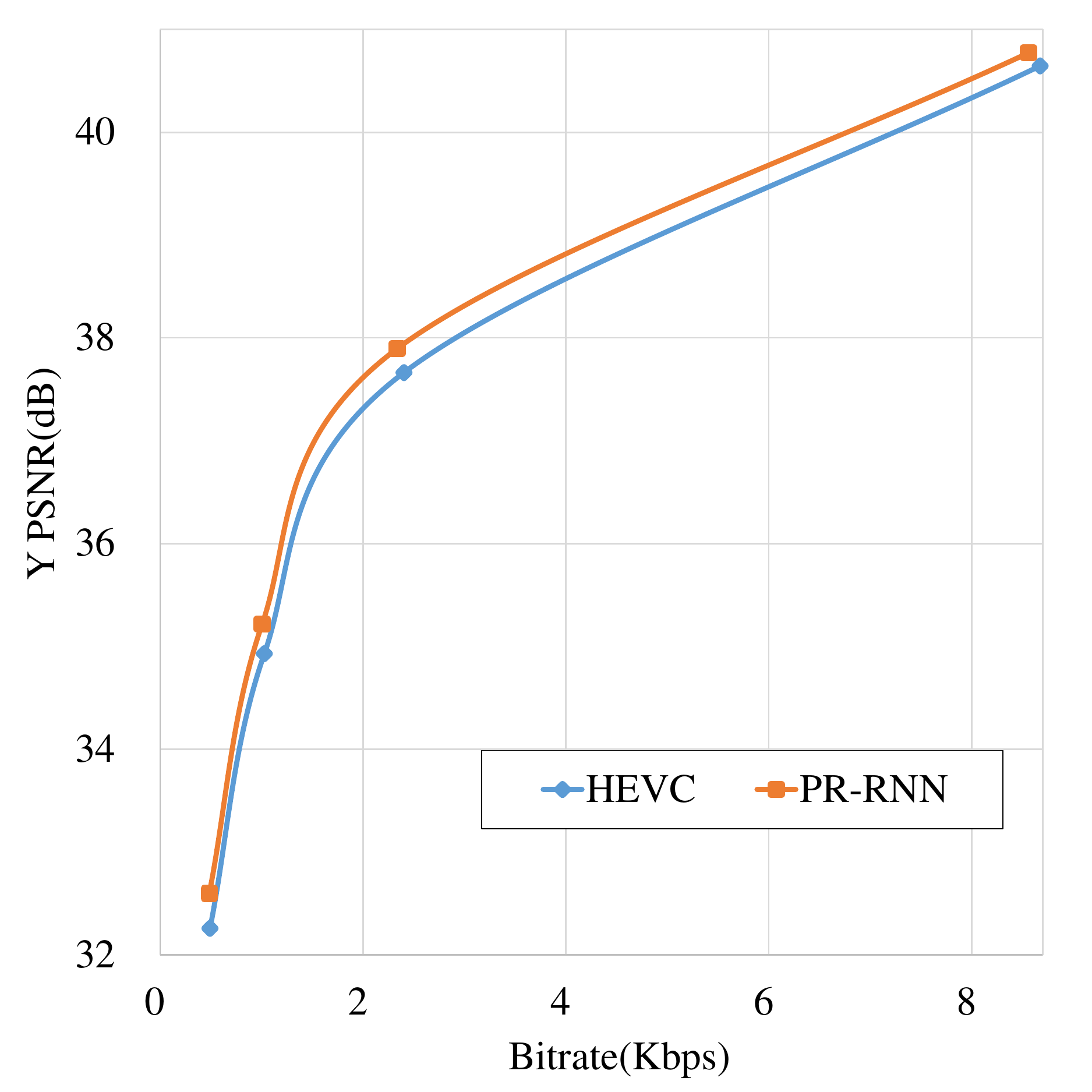}}
\subfigure[Low-Delay P]{
    \includegraphics[width=0.23\linewidth]{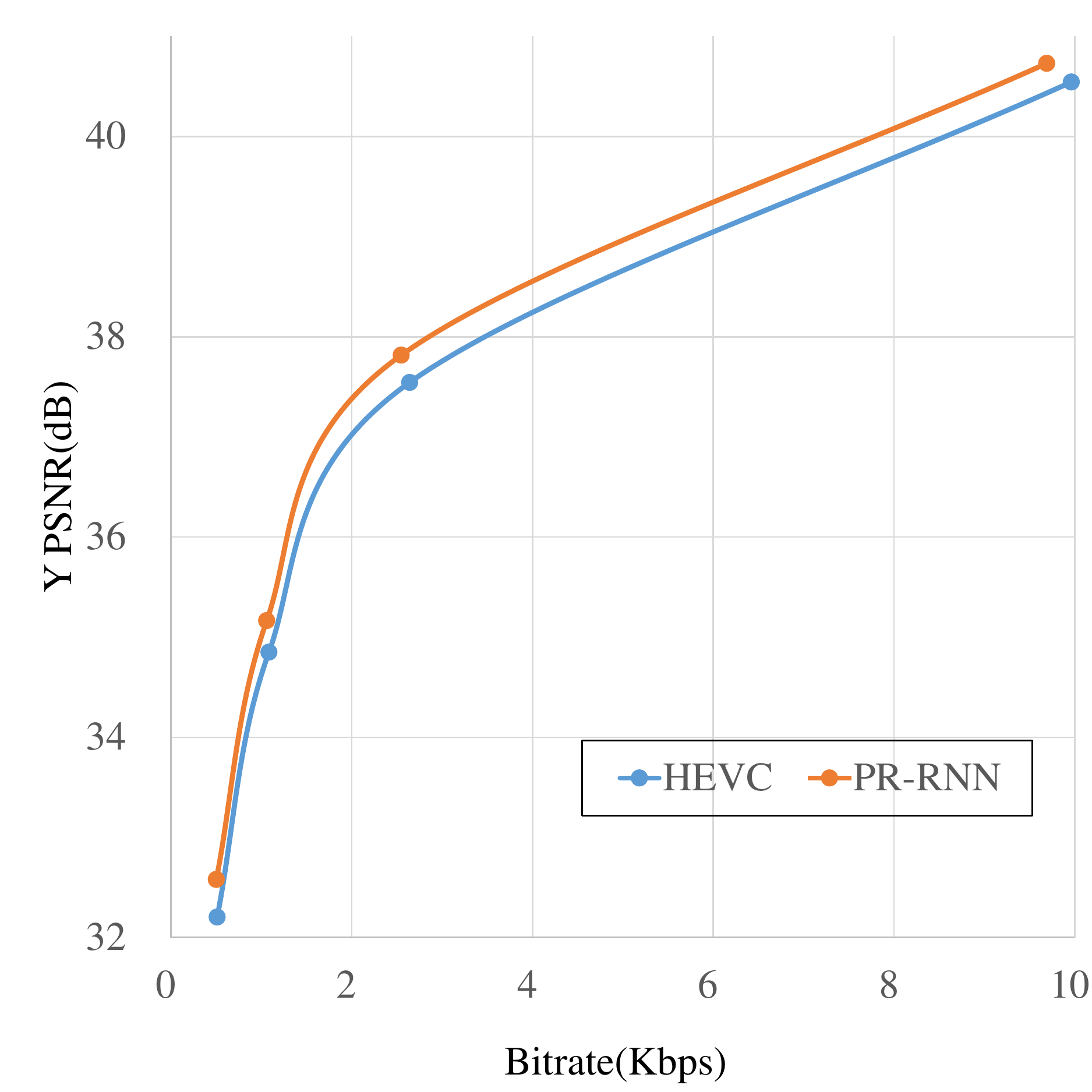}}
\subfigure[Random Access]{
    \includegraphics[width=0.23\linewidth]{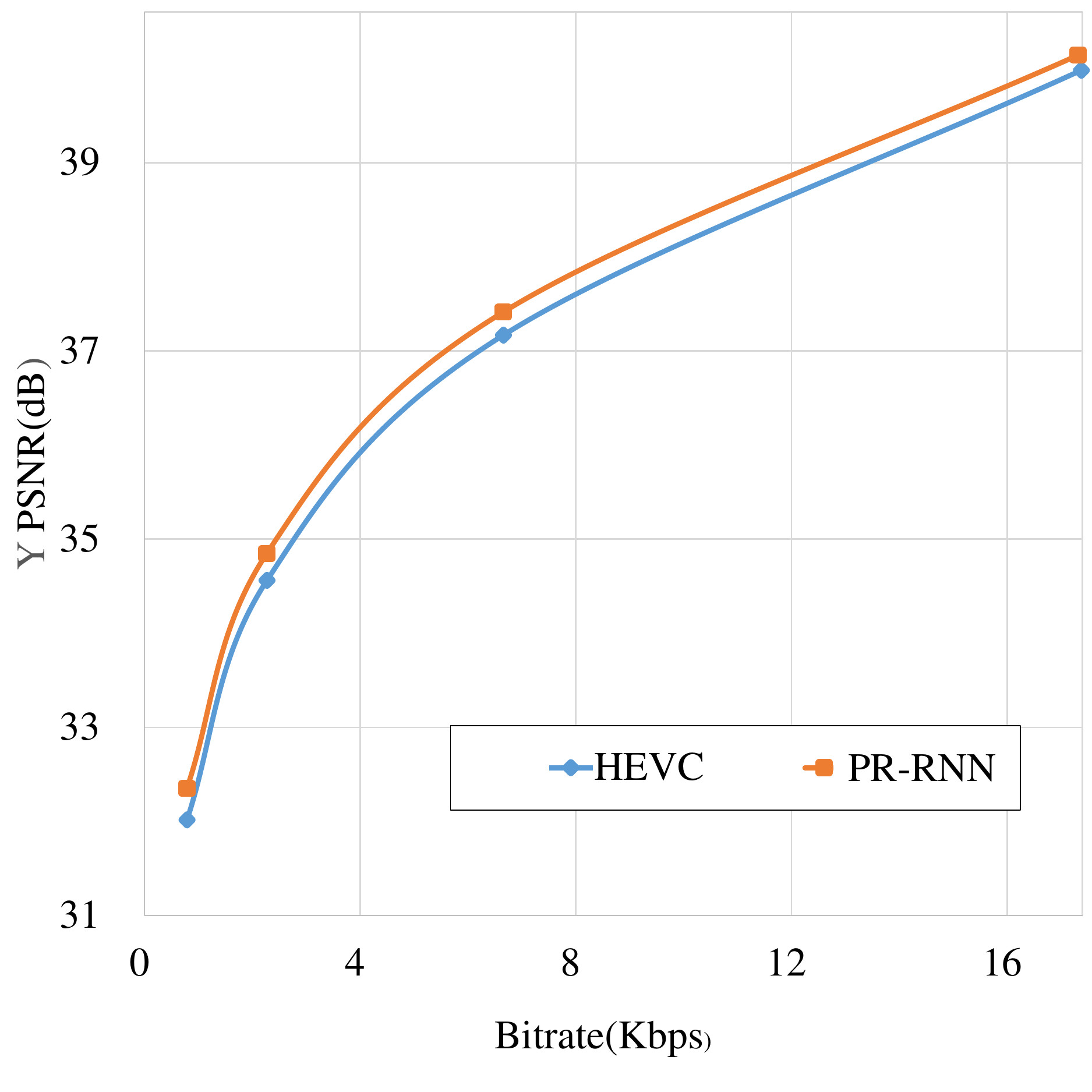}}
	\caption{R-D curves of the test sequences under AI, LDP, LDB and RA configuration.}
	\label{fig:rdcurve}
\end{figure*}

Table \ref{tab:overall} shows the overall performance of our proposed method on classes A, B, C, D and E. Our method has obtained on average $9.0\%$, $9.0\%$, $10.6\%$, $8.0\%$ BD-rate savings, respectively under AI, LDB, LDP and RA configurations. For the
test sequence \textit{Johnny}, up to $17.6\%$ BD-rate saving is obtained for the luma component under LDP configuration. For further verification, we provide rate-distortion (R-D) curves under four configurations as shown in Fig. \ref{fig:rdcurve}. It can be seen that our method is superior to HEVC at all QPs. More significant superiorities are observed especially at higher QP points.


\subsection{Comparison with Existing Methods}
\begin{table}\scriptsize
\centering
\caption{Comparing our PR-CNN with other existing methods under AI configuration.}
\label{tab:AI_comparison}
\begin{tabular}{c|c|c|c|c|c}
    \hline
    \textbf{Class} & \textbf{VRCNN \cite{vrcnn}} & \textbf{DCAD \cite{dcad}} & \textbf{DRN \cite{drn}} & \textbf{RRCNN \cite{RRCNN}} & \textbf{PR-CNN} \\
    \hline
    \hline
    \textbf{A} & -3.5\% & -     & -4.5\% & -9.3\% & \textbf{-10.3\%} \bigstrut[t]\\
\textbf{B} & -3.3\% & -3.4\% & -3.8\% & -6.2\% & \textbf{-6.6\%} \\
\textbf{C} & -5.0\% & -4.6\% & -7.5\% & -9.3\% & \textbf{-10.7\%} \\
\textbf{D} & -5.4\% & -5.2\% & -7.3\% & -9.3\% & \textbf{-9.6\%} \\
\textbf{E} & -6.5\% & -7.8\% & -10.7\% & -11.8\% & \textbf{-13.3\%} \bigstrut[b]\\
\hline
\hline
\textbf{ALL} & -4.6\% & -5.0\% & -6.4\% & -8.7\% & \textbf{-9.7\%} \bigstrut\\
\hline
    \hline
    \end{tabular}%

\end{table}

\begin{table}
    \centering
    
    \caption{Comparing our PR-RNN with other existing methods under LDP configuration.}
    
    \begin{tabular}{c|c|c|c|c}
    \hline
    \textbf{Class} & \textbf{Non-local \cite{non-local}} & \textbf{RHCNN \cite{RHCNN}} & \textbf{MIF \cite{MIF}} & \textbf{PR-RNN} \\
    \hline
    \hline
    \textbf{B} & -23.9\% & -20.8\% & -24.4\% & \textbf{-28.6\%} \\
    \textbf{C} & -13.7\% & -14.5\% & -17.0\% & \textbf{-21.1\%} \\
    \textbf{D} & -8.8\% & -9.8\% & -11.9\% & \textbf{-15.7\%} \\
    \textbf{E} & -27.6\% & -27.3\% & -32.2\% & \textbf{-36.2\%} \\
    \hline
    \hline
    \textbf{ALL} & -18.3\% & -17.7\% & -20.9\% & \textbf{-25.4}\% \\
    \hline
    \end{tabular}%

\label{tab:LD_comparison}
\end{table}

\begin{table}
    \centering

    \caption{Comparing our PR-RNN with other existing methods under RA configuration.}
 \begin{tabular}{c|c|c|c|c}
    \hline
    \textbf{Class} & \textbf{Non-local \cite{non-local}} & \textbf{RHCNN \cite{RHCNN}} & \textbf{MIF \cite{MIF}} & \textbf{PR-RNN} \\
    \hline
    \hline
    \textbf{A} & -9.1\% & -11.2\% & -14.5\% & {\textbf{-15.1\%}} \bigstrut[t]\\
\textbf{B} & -8.5\% & -10.9\% & -13.6\% & \textbf{-14.5\%} \\
\textbf{C} & -4.5\% & -7.1\% & -9.1\% & \textbf{-12.1\%} \\
\textbf{D} & -2.5\% & -5.7\% & -7.3\% & \textbf{-9.0\%} \\
\textbf{E} & -8.2\% & -12.7\% & -15.4\% & \textbf{-19.9\%} \bigstrut[b]\\
\hline
\hline
\textbf{ALL} & -6.3\% & -9.2\% & -11.6\% & \textbf{-13.7\%} \bigstrut\\
    \hline
    \end{tabular}%

    \label{tab:RA_comparison}
\end{table}

Furthermore, we compare our method with some state-of-the-art methods under AI, LDP and RA configurations. The tested sequences are from Class B to Class E. The results are shown in Table \ref{tab:AI_comparison}, \ref{tab:LD_comparison} and \ref{tab:RA_comparison} respectively to validate the superiority of our PRNs.

Under AI configuration, we choose VRCNN \cite{vrcnn} and DCAD \cite{dcad} targeting at post-processing instead of in-loop filtering for comparison. However, as all frames are encoded with no reference frames available during the coding process, the comparison is quite fair. DRN \cite{drn} and RRCNN \cite{RRCNN} are recently proposed in-loop filtering methods for intra frames. DRN consists of Dense Residue Units (DRU) while RRCNN is a recursive residual convolutional neural network. It can be observed that, our PR-CNN outperforms all three compared methods.

When the inter prediction is available, PR-RNN is used to  collaborate with PR-CNN to exploit temporal redundancy. Here, we choose Non-local \cite{non-local}, RHCNN \cite{RHCNN} and MIF \cite{MIF} as the compared methods. In \cite{MIF}, Li \textit{et al.} implemented all three methods on HM 16.5. To make it fair, we also implement our networks based on HM 16.5 and test on class B to class E to show the superiority of our method. It is noted that the compared anchor is HM 16.5 without DF and SAO. It can be observed that our PR-RNN outperforms all three methods under LDP and RA configurations in all tested classes.
When it comes to the LDB configuration, the condition is quite similar to the LDP configuration. Here, to make the experiment more complete, we also compare our method with MIF \cite{MIF} under the LDB configuration. Our method outperforms MIF by 2.5\%, 5.9\%, 4.9\% and 5.1\% on Class B to E, respectively. To make the comparison fair, the model we use is trained by sequences compressed under LDP configuration, which is same as MIF.

\subsection{Ablation Study}
We also conduct some ablation study to verify the necessities and rationality of our design.

\begin{table}[tb]
    \centering
    \caption{The BD-rate results and encoding time of RDN \cite{rdn}, PR-CNN-B under AI configuration.}
   
\begin{tabular}{c|c|c}
    \hline
    \textbf{Class} & \textbf{RDN \cite{rdn}} & \multicolumn{1}{c}{\textbf{PR-CNN-B}} \\
    \hline
    \hline
    \textbf{B} & -5.7\% & \textbf{-6.0\%} \\
    \textbf{C} & -9.7\% & \textbf{-10.1\%} \\
    \textbf{D} & -8.9\% & \textbf{-9.2\%} \\
    \textbf{E} & -12.0\% & \textbf{-12.5\%} \\
    \hline
    \hline
    \textbf{ALL} & -8.7\% & \textbf{-9.1\%} \\
    \hline
    \textbf{Time} & 152.2\% & 153.6\% \\
    \hline
    \end{tabular}%
    \label{tab:PR-CNN}
\end{table}

\begin{table}[htbp]
  \centering
  \caption{BD-rate results of PR-CNN baseline, PR-CNN w/ SM-CU and PR-CNN w/ MM-CU. Separate and Non-Separate denote two fusion ways. }

    \begin{tabular}{c|c|c|c|c}
    \hline
    \multirow{2}[2]{*}{\textbf{Class}} & \multicolumn{1}{c|}{\multirow{2}[2]{*}{\textbf{PR-CNN-B}}} & \multicolumn{1}{c|}{\multirow{2}[2]{*}{\textbf{SM-CU}}} & \multicolumn{2}{c}{\textbf{MM-CU}} \\
          &       &       & \multicolumn{1}{p{6em}}{\textbf{Non-Separate}} & \textbf{Separate} \\
    \hline
    \hline
    \textbf{B} & -6.0\% & -6.2\% & -6.4\% & \textbf{-6.6\%} \\
    \textbf{C} & -10.1\% & -10.4\% & -10.6\% & \textbf{-10.7\%} \\
    \textbf{D} & -9.2\% & -9.3\% & -9.2\% & \textbf{-9.6\%} \\
    \textbf{E} & -12.5\% & -13.0\% & -13.2\% & \textbf{-13.3\%} \\
    \hline
    \hline
    \textbf{ALL} & -9.1\% & -9.3\% & -9.4\% & \textbf{-9.6\%} \\
    \hline
    \end{tabular}%
  \label{tab:MM-CU}%
\end{table}%

\textbf{Verification of Inter-Block Connection.}
To verify the superiority of our PRB, we train an RDN \cite{rdn} and a PR-CNN \textbf{B}aseline model (denoted as PR-CNN-B) which only takes the unfiltered frame as input without side information. They are trained by the same training procedure as we mentioned before. We test the BD-rate under AI configuration. As shown in Table \ref{tab:PR-CNN}, original RDN provides a 8.7\% performance gain. An additional inter-block connection can further enhance the coding performance by a margin of 0.4\%. Further, in Table \ref{tab:PR-CNN}, encoding time comparison is also provided. The time is calculated based on HM 16.15 anchor. The result shows that, almost no additional time cost is paid to improve the overall performance by introducing the inter-block connection.

\textbf{Verification of MM-CU Maps.}
To verify the effectiveness of our proposed MM-CU Maps, we conduct an experiment to compare the performance of PR-CNN full model (\textit{i.e.} with MM-CU as input) to PR-CNN \textbf{B}aseline (PR-CNN-B) model and PR-CNN-B with Single scale M-CU map (SM-CU) guidance. Besides, we compare two fusion ways: (1) concatenating all MM-CU maps together to extract features and then fuse it into the baseline network, which is denoted as Non-Separate in Table \ref{tab:MM-CU}; (2) our proposed fusion mechanism inserting different maps into different depths, denoted as Separate. 
As it is shown in Table \ref{tab:MM-CU}, compared to PR-CNN-B and SM-CU, the performance of PR-CNN with MM-CU maps guidance further increases by 0.5\% and 0.3\% under AI configuration. In addition, the fusion way plays a role in the overall performance which improves the BD-rate by 0.2\%.

\begin{table}[tb]
    \centering
    
    \caption{Ablation study to verify the effectiveness of the Collaborative Mechanism. The experiment is implemented under LDP configuration.}
\begin{tabular}{c|c|c|c}
\hline
\textbf{Class} & \textbf{Sequence} & \textbf{PRN+WarpN} & \textbf{PR-RNN-N} \bigstrut\\
\hline
\hline
\multirow{5}[4]{*}{\textbf{C}} & BasketballDrill & -11.6\% & \textbf{-12.1\%} \bigstrut[t]\\
      & BQMall & -9.6\% & \textbf{-10.5\%} \\
      & PartyScene & -6.3\% & \textbf{-7.0\%} \\
      & RaceHorsesC & -8.2\% & \textbf{-9.4\%} \bigstrut[b]\\
\cline{2-4}      & Average & -8.9\% & \textbf{-9.7\%} \bigstrut\\
\hline
\hline
\multirow{5}[4]{*}{\textbf{D}} & BasketballPass & -8.3\% & \textbf{-8.8\%} \bigstrut[t]\\
      & BlowingBubbles & -4.5\% & \textbf{-5.2\%} \\
      & BQSquare & -7.4\% & \textbf{-8.1\%} \\
      & RaceHorses & -8.1\% & \textbf{-9.4\%} \bigstrut[b]\\
\cline{2-4}      & Average & -7.1\% & \textbf{-7.9\%} \bigstrut\\
\hline
\hline
\textbf{ALL} & \textbf{Average} & -8.0\% & \textbf{-8.8\%} \bigstrut\\
\hline
\end{tabular}%

    \label{tab:collaborative mechanism}
\end{table}

\textbf{Verification of Collaborative Learning Mechanism.}
Our PR-RNN applies a collaborative learning mechanism to transfer information between states. To verify the effectiveness of collaborative learning mechanism, we train a model which simply aligns the neighbouring frames to current frame by optical flow and concatenates them as the input of a CNN with same PRBs as PR-RNN. The comparison between PRN-Warp and PR-RNN-N can validate the effectiveness collaborative learning mechanism. The result is also shown in Table \ref{tab:collaborative mechanism}. From the last two columns, we can find that, PR-RNN-N performs better than PRN+WarpN, which demonstrates that our collaborative mechanism method indeed benefits the restoration.

\begin{figure*}
    \centering
    \includegraphics[width=\linewidth]{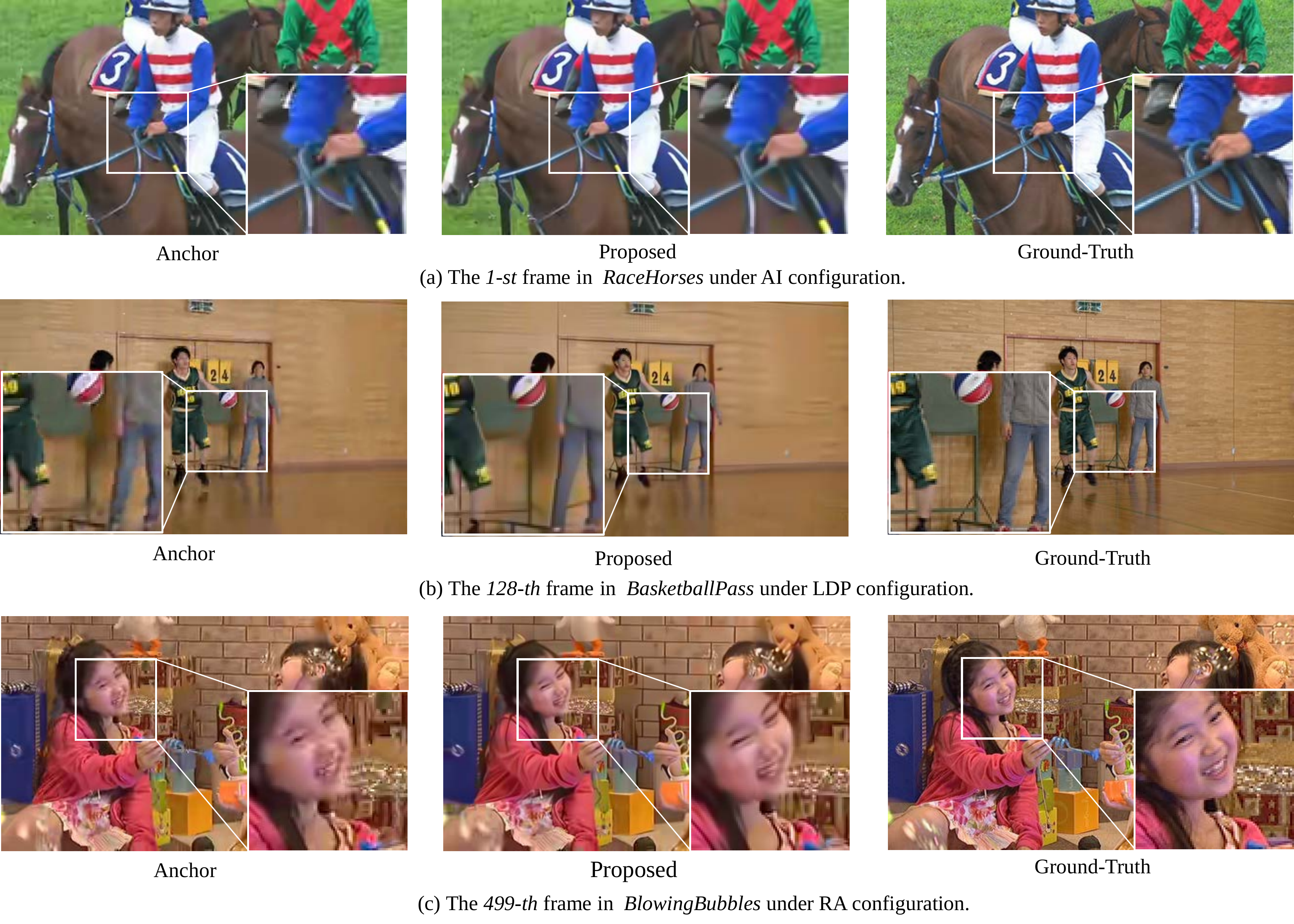}
    \caption{Some subjective results of our proposed method compared to HEVC anchor.}
    \label{fig:subjective}
\end{figure*}

\subsection{Further Analysis on Integration}
\textbf{Generalization Capacity on QP.} As we mentioned before, we train models at 4 QP points, \textit{i.e.} 22, 27, 32, 37. If we need to test other QPs, we can simply reuse the model trained under the closest QP which can avoid the space overcommit. We test the PR-CNN at {21, 26, 31, 36} and {23, 28, 33, 38} with models trained at {22, 27, 32, 37} under AI configuration. We set the coding time to 1 second for each sequence. The results are shown in Table \ref{tab:qp}. According to the results, we can find that, the results vary slightly with respect to QP jittering.

\textbf{Necessity of DF and SAO.} 
In our integration, we insert our model between DF and SAO with both filters on. However, one may wonder that DF and SAO could degrade the overall performance as they would waste bit-stream and improve little quality. To verify the necessity of turning on DF and SAO, we conduct the experiment to compare the performance of PR-CNN with DF and SAO off. We first turn DF and SAO off, then only turn on DF and at last turn both of them on. The experimental results are shown in Table \ref{tab:sao}. From the table, we can find that DF and SAO can brings about 1\% performance gain. Here, the coding time is also set to 1 second for each sequence to accelerate the experiment.

\textbf{Complexity Analysis.} 
We integrate our models into HM with the help of LibTorch.
The number of parameters of our proposed PR-CNN and PR-RNN is 7.24M and 5.03M respectively, which occupies 29.8M and 22.1M disk space. The peak running memory of GPU is about 8.5G.
We compare the encoding and decoding time of our approach with HM anchor under AI and RA configurations. The result is shown in Table \ref{tab:complexity}. In Table \ref{tab:complexity} (a), we also test the runtime of RDN and PR-CNN-B under AI configuration, to show the complexity increase of additionally using the inter-block connection and MM-CU maps. It can be found that, the inter-block connection brings 1.4\% and 23.1\% additional time consumptions in encoding and decoding, respectively, while the MM-CU maps bring 39.3\% and 2,501.6\% runtime increase in encoding and decoding, respectively. In Table \ref{tab:complexity} (b), beyond HM anchor, we further compare our complexity and RD performance with MIF \cite{MIF} under RA configuration. It is observed that our method can obtain a 25.4\% performance gain over HM in spite that the encoding time nearly doubles on GPU. Compared to MIF, our method obtains a 5.7\% performance gain with little increase in encoding and decoding time. Here we also provide the time complexity of PR-RNN on CPU, where the speed drops a lot compared to that on GPU.

\begin{table}[htbp]
  \centering
  \caption{Evaluation On Generalization Capacity on QP.}
    \begin{tabular}{c|c|c|c}
    \hline
    \textbf{Class} & \textbf{Original QP} & \multicolumn{1}{c|}{\textbf{QP+1}} & \multicolumn{1}{c}{\textbf{QP-1}} \\
    \hline
    \hline
    \textbf{B} & -6.9\% & -6.5\% & -5.9\% \\
    \textbf{C} & -8.6\% & -8.0\% & -7.2\% \\
    \textbf{D} & -6.7\% & -6.1\% & -5.4\% \\
    \textbf{E} & -13.2\% & -12.7\% & -11.8\% \\
    \hline
    \hline
    \textbf{ALL} & -8.7\% & -8.1\% & -7.3\% \\
    \hline
    \end{tabular}%
  \label{tab:qp}%
\end{table}%

\begin{table}[htbp]
  \centering
  \caption{Necessity of DF and SAO.}
    \begin{tabular}{c|c|c|c}
    \hline
    \textbf{Class} & \textbf{w/o DF \& SAO} & \textbf{w/o SAO} & \multicolumn{1}{c}{\textbf{Full}} \\
    \hline
    \hline
    \textbf{B} & -5.2\% & -6.5\% & \textbf{-6.9\%} \\
    \textbf{C} & -8.1\% & -8.3\% & \textbf{-8.6\%} \\
    \textbf{D} & -5.9\% & -6.2\% & \textbf{-6.7\%} \\
    \textbf{E} & -12.3\% & -13.0\% & \textbf{-13.2\%} \\
    \hline
    \hline
    \textbf{ALL} & -7.7\% & -8.3\% & \textbf{-8.7\%} \\
    \hline
    \end{tabular}%
  \label{tab:sao}%
\end{table}%

\begin{table}[htbp]
  \centering
  \caption{Encoding and decoding runtime}
    \subtable[Runtime under AI configuration]{
    
        \begin{tabular}{p{7.25em}|c|c|c}
\hline
\textbf{Model} & \multicolumn{1}{p{4.835em}|}{\textbf{RDN [13]}} & \multicolumn{1}{p{6.085em}|}{\textbf{PR-CNN-B}} & \multicolumn{1}{p{4.585em}}{\textbf{PR-CNN}} \bigstrut\\
\hline
\hline
\textbf{Encoding Time} & 152.2\% & 153.6\% & 192.9\% \bigstrut[t]\\
\textbf{Decoding Time} & 3,329.8\% & 3,306.7\% & 5,808.3\% \bigstrut[b]\\
\hline
\end{tabular}%
    }
    
    \subtable[Runtime under RA configuration]{
        \begin{tabular}{l|cc|c}
\hline
\textbf{Anchor} & \multicolumn{2}{c|}{\textbf{HM}} & \textbf{MIF \cite{MIF}} \bigstrut\\

\textbf{Device} & \textbf{CPU} & \textbf{GPU} & \textbf{GPU} \bigstrut\\
\hline
\hline
\textbf{Encoding Time} & 8,162\% & 196\% & 105\% \bigstrut[t]\\
\textbf{Decoding Time} & 2,395,126\% & 28,652\% & 123\% \bigstrut[b]\\
\hline
\textbf{Performance Gain} & \multicolumn{2}{c|}{-25.4\%} & -5.7\% \bigstrut\\
\hline
\end{tabular}%

    }
    
  \label{tab:complexity}%
\end{table}%

\subsection{Subjective Results}
We compare the subjective quality of HEVC anchor and our proposed method. Fig. \ref{fig:subjective} illustrates the some examples which are compressed under AI, LDP and RA configurations respectively when QP is 37. For \textit{RaceHorses}, it can be observed that the rein is blurry in the results of the HM anchor but becomes more clear after being filtered by our proposed method. In \textit{BasketballPass}, the bottom of the gate is missed in the results of HM anchor but appears in that of our PR-RNN. For \textit{BlowingBubbles}, the girl's face is degraded by multiple artifacts and our filtered result shows better visual quality. All these examples show that, our approach is superior to HEVC in subjective visual qualities.

\section{Conclusion}
\label{sec6}
In this paper, we propose Progressive Rethinking Networks with Collaborative Learning Mechanism. We design a Progressive Rethinking Block to introduce inter-block connections to compensate for possible information lost across blocks. Furthermore, we extract side information from HEVC codecs to facilitate restoration. We generate Multi-scale Mean value of Coding Unit maps by calculating the mean value of the CU each time a partition happens and replacing the original pixel value with the mean value. The MM-CU maps are fused to a convolutional neural network consisting of PRBs called Progressive Rethinking Convolutional Neural Network. Beyond that, we develop a Collaborative Learning Mechanism to effectively utilize temporal side information. In our collaborative learning mechanism, not only the state of current frame but also the states of reference frames are updated. We implement our collaborative learning mechanism through a Recurrent Neural Network called PR-RNN. Experimental results show that our PR-CNN outperforms HEVC baseline by 9.0\% and PR-RNN outperforms HEVC baseline by 9.0\%, 10.6\% and 8.0\% under LDB, LDP and RA configurations.

\begin{IEEEbiography}[{\includegraphics[width=1in,height=1.25in,keepaspectratio]{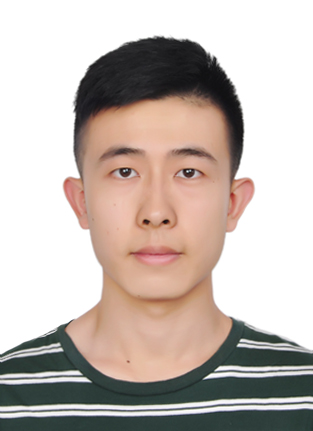}}]
    {Dezhao Wang} (STM'19) received the B.S. degree in computer science fom Peking University, Beijing, China in 2020, where he is currently
working toward the master's degree with Wangxuan
Institute of Computer Technology, Peking
University. His current research interests include
video and image compression.
\end{IEEEbiography}

\begin{IEEEbiography}[{\includegraphics[width=1in,height=1.25in,keepaspectratio]{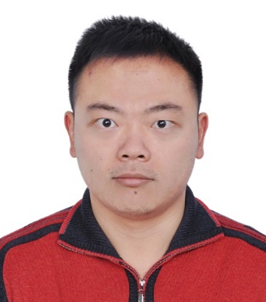}}]
    {Sifeng Xia} received the B.S. and Master degrees in computer science from Peking University, Beijing, China, in 2017 and 2020, respectively. His current research interests include deep learning-based image processing and video coding.
\end{IEEEbiography}

\begin{IEEEbiography}[{\includegraphics[width=1in,height=1.25in,keepaspectratio]{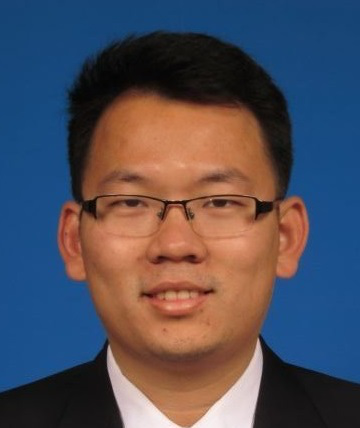}}]
	{Wenhan Yang} (M'18) 
	received the B.S degree and Ph.D. degree (Hons.) in computer science from Peking University, Beijing, China, in 2012 and 2018. He is currently a postdoctoral research fellow with the Department of Computer Science, City University of Hong Kong. Dr. His current research interests include image/video processing/restoration, bad weather restoration, human-machine collaborative coding. He has authored over 100 technical articles in refereed journals and proceedings, and holds 9 granted patents.
	He received the IEEE ICME-2020 Best Paper Award, the IFTC 2017 Best Paper Award, and the IEEE CVPR-2018 UG2 Challenge First Runner-up Award.
	He was the Candidate of CSIG Best Doctoral Dissertation Award in 2019. He served as the Area Chair of IEEE ICME-2021, and the Organizer of IEEE CVPR-2019/2020/2021 UG2+ Challenge and Workshop.
\end{IEEEbiography}

\begin{IEEEbiography}[{\includegraphics[width=1in,height=1.25in,clip,keepaspectratio]{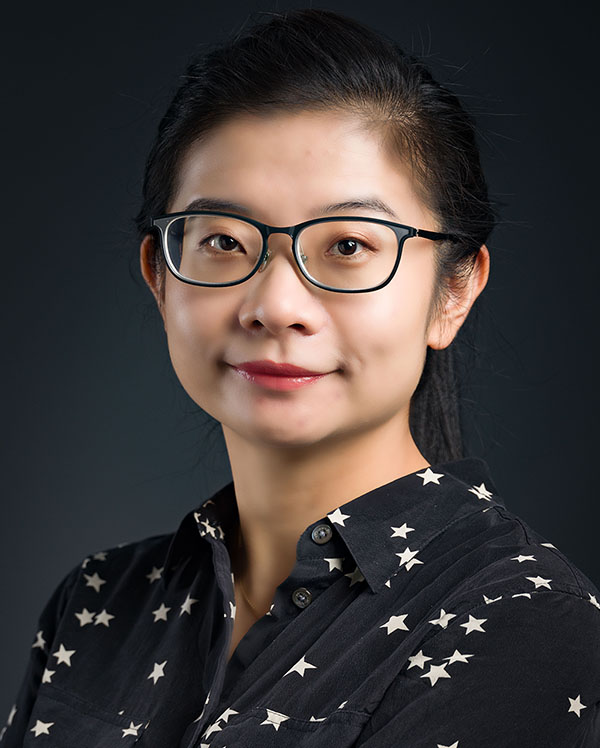}}]
	{Jiaying Liu} (M'10-SM'17) 
	is currently an Associate Professor, Peking University Boya Young Fellow with the Wangxuan Institute of Computer Technology, Peking University. She received the Ph.D. degree (Hons.) in computer science from Peking University, Beijing China, 2010. She has authored over 100 technical articles in refereed journals and proceedings, and holds 50 granted patents. Her current research interests include multimedia signal processing, compression, and computer vision. 
	
	Dr. Liu is a Senior Member of IEEE, CSIG and CCF. She was a Visiting Scholar with the University of Southern California, Los Angeles, from 2007 to 2008. She was a Visiting Researcher with the Microsoft Research Asia in 2015 supported by the Star Track Young Faculties Award. She has served as a member of Multimedia Systems \& Applications Technical Committee (MSA TC), and Visual Signal Processing and Communications Technical Committee (VSPC TC) in IEEE Circuits and Systems Society. She received the IEEE ICME-2020 Best Paper Award and IEEE MMSP-2015 Top10\% Paper Award. She has also served as the Associate Editor of IEEE Trans. on Image Processing, IEEE Trans. on Circuit System for Video Technology and Elsevier JVCI, the Technical Program Chair of IEEE ICME-2021/ACM ICMR-2021, the Publicity Chair of IEEE ICME-2020/ICIP-2019, and the Area Chair of CVPR-2021/ECCV-2020/ICCV-2019. She was the APSIPA Distinguished Lecturer (2016-2017).
\end{IEEEbiography}

\bibliographystyle{IEEEtran}
\bibliography{refs}

\end{document}